\documentclass[reprint,amsmath,amssymb,prb,aps,showpacs]{revtex4-1}
\usepackage{hyperref}
\usepackage{bbm}
\usepackage{tikz}
\usepackage[vcentermath]{youngtab}

\newcommand{\includediagram}[2]{
\begin{array}{c}
\includegraphics[height=#1]{#2}
\end{array}
}

\DeclareMathOperator{\tr}{\text{tr}}
\newcommand{\mat}{\begin{pmatrix}}
\newcommand{\tam}{\end{pmatrix}}
\newcommand{\smat}{\left(\begin{smallmatrix}}
\newcommand{\stam}{\end{smallmatrix}\right)}

\newcommand{\tinyyoung}[1]{{\tiny\young(#1)}}

\newcommand{\tinyyng}[1]{{\tiny\yng(#1)}}

\newcommand{\idop}{\mathbbm{1}}
\newcommand{\bC}{\mathbb{C}}
\newcommand{\bK}{\mathbb{K}}
\newcommand{\bP}{\mathbb{P}}
\newcommand{\bQ}{\mathbb{Q}}
\newcommand{\bX}{\mathbb{X}}
\newcommand{\Integer}{\mathbb{Z}}
\newcommand{\cN}{\mathcal{N}}
\newcommand{\cP}{\mathcal{P}}
\newcommand{\cV}{\mathcal{V}}

\begin{document}

\title{Chiral Haldane phases of SU(N) quantum spin chains in the
  adjoint representation}

\author{Abhishek Roy}
\email{aroy@thp.uni-koeln.de}
\author{Thomas Quella}%
\email{Thomas.Quella@uni-koeln.de}
\affiliation{Institute of Theoretical Physics, University of Cologne\\
  Z\"ulpicher Stra\ss{}e 77, D-50937 Cologne, Germany }%
\date{\today}
\begin{abstract}
  Gapped quantum spin chains with symmetry
  $\text{PSU(N)}=\text{SU(N)}/\Integer_N$ are known to possess $N$
  distinct symmetry protected topological phases. Besides the trivial
  phase, there are $N-1$ Haldane phases which are distinguished by the
  occurrence of massless boundary spins. Motivated by the potential
  realization in alkaline-earth atomic Fermi gases, we explicitly
  construct previously unknown Hamiltonians for two classes of chiral
  AKLT states and we discuss their physical properties. We also point
  out a deep connection between symmetry protection in gapped and
  gapless 1D quantum spin systems and its implications for a potential
  multicritical nature of topological phase transitions.
\end{abstract}

\pacs{03.65.Vf, 75.10.Pq, 75.10.Kt}
\maketitle

\section{\label{sc:Introduction}Introduction}

Historically, the investigation of SU(2) quantum spin chains has been
a great source of inspiration for theoretical physics. Besides Bethe's
solution of the $S=1/2$ spin chain, one of the most notable
achievements in this context was Haldane's discovery that the physical
properties of the SU(2) Heisenberg model in one dimension depend
crucially on the type of spin that is used.\cite{Haldane:1983464} For
half-integer spins the system can easily be shown to be
gapless\cite{Lieb:1961fr,Affleck:1986pq} while for integer spins it
has been conjectured to develop a gap, the so-called Haldane gap,
while preserving translation invariance.\cite{Haldane:1983464} This
conjecture triggered a tremendous amount of work before it could be
confirmed for various values of the spin using analytical and
numerical approaches, or even in
experiment.\cite{Haldane:1983464,Affleck:1987PhRvB..36.5291A,Affleck:1989JPCM....1.3047A,Granroth:1996PhRvL..77.1616G,Wang:1999PhRvB..6014529W}

The gapped phase of SU(2) spin chains with unique translation
invariant ground state which arises for integer spin representations
is nowadays commonly called the Haldane phase. As has been understood
only recently, it provides a paradigmatic example of a symmetry
protected topological phase.\cite{Pollmann:2012PhRvB..85g5125P} This
means that its ground state cannot be transformed into a trivial
product state except if either the relevant symmetry is broken or the
system undergoes a phase transition. The symmetries that qualify for
the protection of the Haldane phase are
$\text{SO(3)}=\text{SU(2)}/\Integer_2$, its dihedral subgroup
$\Integer_2\times\Integer_2$, time-reversal or space inversion. The
prototypical example for a Hamiltonian and a ground state which reside
in the Haldane phase is provided by the AKLT construction, named after
its protagonists Affleck, Kennedy, Lieb and
Tasaki.\cite{Affleck:PhysRevLett.59.799,Affleck:1987cy} It provides
the first realization of what these days is commonly referred to as a
matrix product state.

The study of Haldane phases reaches a new level of complexity for spin
systems with higher rank symmetries such as SU(N). In contrast to
SU(2) there are now up to $N-1$ Haldane
phases.\cite{Duivenvoorden:2012arXiv1206.2462D} In each of them the
ground state, which is assumed to be unique, is characterized by a
distinct $\Integer_N$-valued non-zero topological quantum number. The
latter can be measured in terms of a non-local string order
parameter\cite{Duivenvoorden:2012arXiv1208.0697D} and it is also
reflected in the nature of emergent fractionalized boundary spins
which arise if the system is studied with open boundary
conditions.\cite{Duivenvoorden:2012arXiv1206.2462D} Information about
the topological class of the system can also be inferred by looking at
certain Berry phases\cite{Motoyama:2015arXiv150800960M} or at the
behavior of specific $\Integer_N\times\Integer_N$ subgroups of
$\text{PSU(N)}=\text{SU(N)}/\Integer_N$.\cite{Duivenvoorden:2013arXiv1304.7234D,Else:2013arXiv1304.0783E}

Since the classification of SU(N) Haldane phases and the proof of
existence in Ref.~\onlinecite{Duivenvoorden:2012arXiv1206.2462D} were
based on abstract principles, the article did not give a detailed
prescription for how to realize them in the phase diagram of concrete
spin models with specific physical SU(N) spins. The first systematic
attempt to overcome this deficiency took place in
Ref.~\onlinecite{Duivenvoorden:2012arXiv1208.0697D} where a
one-dimensional cut through the phase diagram of a SU(3)-invariant
system with spins transforming in the eight-dimensional adjoint
representation was studied. Along this line of interpolation a
topological phase transition was observed which connects the two
distinct SU(3) Haldane phases with topological quantum number
$\pm[1]\in\Integer_3$. The phase diagram of this model was then
further explored in Ref.~\onlinecite{Morimoto:2014PhRvB..90w5111M}.

Another concrete proposal, now with SU(N) spins transforming in the
so-called ``self-conjugate'' representation, was provided in
Ref.~\onlinecite{Nonne:2012arXiv1210.2072N}. The definition of this
representation is bound to even values of $N$ and the associated
Haldane phase is characterized by the topological quantum number
$[\tfrac{N}{2}]\in\Integer_N$. Further investigations illuminated
various complementary aspects of this
model.\cite{Bois:2015PhRvB..91g5121B,Tanimoto:2015arXiv150807601T,Quella:2015}
However, what is still lacking to date is a systematic construction of
SU(N) Haldane phases which is valid for general values of $N$.

The main goal of our current paper is to close this gap and to provide
explicit realizations of the four elementary non-trivial Haldane
phases with topological quantum numbers $\pm[1]$ and $\pm[2]$
(regarded as elements of $\Integer_N$). Following the ideas of
Ref.~\onlinecite{Duivenvoorden:2012arXiv1208.0697D} we choose to work
with physical SU(N) spins transforming in the adjoint
representation. We then construct AKLT states whose auxiliary spins
transform in either the (anti-)fundamental or the rank-2
anti-symmetric representation and we determine the associated parent
Hamiltonians. Moreover, we also confirm the existence of a mass
gap. For the special case SU(4), our construction leads to a
realization of the complete set of three Haldane phases in a single
phase diagram.

AKLT models for SU(N) attracted some
attention\cite{Affleck:1991NuPhB.366..467A,Greiter:PhysRevB.75.184441,Rachel:2010JPhCS.200b2049R}
even before realizing the intimate connection to the developing field
of symmetry protected topological phases.\footnote{It should be
  emphasized that not every AKLT or matrix product state is
  automatically symmetry protected. A prominent example is the SU(2)
  AKLT state for spin $S=2$ which is based on auxiliary $S=1$ spins.}
Also our current setup with an AKLT state based on the adjoint
representation and with auxiliary spins in the (anti-)fundamental
representation already received a thorough treatment in the
past. However, the authors of
Ref.~\onlinecite{Affleck:1991NuPhB.366..467A,Greiter:PhysRevB.75.184441,Rachel:2010JPhCS.200b2049R}
(and later also those of
Ref.~\onlinecite{Morimoto:2014PhRvB..90w5111M}) have been satisfied
with writing down a Hamiltonian which exhibits spontaneous inversion
symmetry breaking. In this case, there are two ground states which
both possess a simple AKLT form. The two states are related by an
exchange of auxiliary spins and correspondingly space inversion maps
one into the other.

From the perspective of the general
classification\cite{Duivenvoorden:2012arXiv1206.2462D} these systems,
however, do not reside in a well-defined pure Haldane phase but rather
in a superposition of two different ones which are distinguished by
their topological quantum numbers. Our current analysis resolves this
degeneracy by incorporating suitable terms in the Hamiltonian which
explicitly break inversion symmetry while preserving SU(N). The precise
form of the desired Hamiltonian is derived using a powerful
diagrammatic method called ``birdtracks''.\cite{Cvitanovic} The same
technique also gives straightforward access to the eigenvalues of the
ground states' transfer matrix which, in turn, determine entanglement
spectrum, spin-spin correlation functions and correlation length.

Another important outcome of our diagrammatic approach is a new type
of universal parent Hamiltonian. While not a projector in general, it
has the great advantage of having an extremely simple graphical
(tensor) form.

Our analysis paves the way to a more detailed study of the phase
diagram of SU(N) spin chains, including topological phase
transitions. According to a recent proposal by Furuya and Oshikawa,
the notion of symmetry protection is not only restricted to gapped
phases but can also be used to characterize critical
points.\cite{Furuya:2015arXiv150307292F} Specifically it leads to
selection rules which impose restrictions to the existence of
potential renormalization group flows between conformal field
theories. While initially analyzed for SU(2), this statement was
recently extended to SU(N) by
Lecheminant.\cite{Lecheminant:2015arXiv150901680L} We will add a
further layer of insight to this analysis by showing that topological
phase transitions between symmetry protected topological phases may
generally be forced to be described by multicritical points, just by
symmetry considerations.

Deeply intertwined with the study of Haldane phases is the
investigation of a potential generalization of Haldane's Conjecture to
SU(N) spin systems, see Ref.~\onlinecite{Affleck:1986pq} for the
original conjecture, Ref.~\onlinecite{Affleck:1989JPCM....1.3047A} for
a review and
Refs.~\onlinecite{Greiter:PhysRevB.75.184441,Rachel2010:PhysRevB.80.180420,Duivenvoorden:2012arXiv1206.2462D}
for more recent studies and proposals. As suggested in
Ref.~\onlinecite{Duivenvoorden:2012arXiv1206.2462D}, the SU(N)
Heisenberg model should generally lead to a gapped phase if and only
if the physical spin permits the construction of a symmetry protected
topological phase. Otherwise it should be gapless. Our current
analysis leads us to our own version of Haldane's Conjecture and to an
educated guess about the nature of the phase realized by the SU(4)
Heisenberg Hamiltonian for adjoint representation.

Last but not least, our analysis should also be seen in connection
with recent experimental progress on the realization of SU(N)
magnetism in ultra-cold gases of fermionic alkaline-earth
atoms.\cite{Gorshkov:2010NatPh...6..289G,Zhang:2014Sci...345.1467Z,Scazza:2014NatPh..10..779S,Pagano:2014NatPh..10..198P,Hofrichter:2015arXiv151107287H}
The adjoint representation we are using to realize our Haldane phases
is considerably smaller than the self-conjugate representation
suggested in Ref.~\onlinecite{Nonne:2012arXiv1210.2072N}. It is
therefore conceivable that it enjoys a better stability when realized
in terms of elementary fermionic degrees of freedom and better
experimental addressability. For a general discussion of
one-dimensional SU(N) spin systems and their realization in the strong
coupling limit of Fermi-Hubbard models we refer to a recent
review.\cite{Capponi:2015arXiv150904597C}.

Our paper is organized as follows. In Section~\ref{sc:Setup} we
introduce the physical setup, i.e.\ we specify the physical SU(N) spin
we would like to study and we classify the terms which may enter a
general SU(N)-invariant Hamiltonian with nearest neighbor
interactions. The technical foundations of the paper are presented in
Section~\ref{sc:Haldane}. After outlining the general philosophy of
the AKLT construction we discuss in detail which auxiliary spins are
permitted if the physical spin is to transform in the adjoint
representation. We then provide an exhaustive discussion of
SU(N)-invariant two-site operators, with a particular focus on higher
order spin-spin couplings. In order to discuss the properties of these
operators we then introduce birdtracks, a convenient graphical
calculus, which provides a simple way to determine their eigenvalues.
Our main results are contained in Section~\ref{sc:Hamiltonians}. Here
we explicitly state the AKLT Hamiltonians which give rise to chiral
Haldane phases with topological quantum number $\pm[1]$ and
$\pm[2]$. This is followed by a brief discussion of the special cases
SU(3) and SU(4).

In Section~\ref{sc:Correlations} we show how the graphical calculus
may be used to determine the eigenvalues of the transfer matrix,
resulting in a proof of the gapfulness of our AKLT states. A more
detailed exploration of the phase diagram is provided in
Section~\ref{sc:PhaseDiagram}, with a particular focus on critical
phases. In this context we also discuss the nature of potential
topological phase transitions and the implications of symmetries for
the occurrence of multicritical points. In
Section~\ref{sc:ParentHamiltonian} we present our idea of a universal
parent Hamiltonian. This is then adapted to compute the precise form
of the AKLT Hamiltonians under consideration. Our approach leads to
some thoughts on Haldane's Conjecture for SU(N) quantum spin models
which are presented in Section \ref{sc:Conjecture}. We conclude our
paper with a brief summary and an outlook to future research.

\section{\label{sc:Setup}Physical setup}

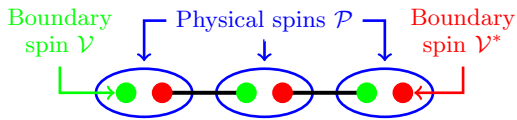
\begin{figure}
\begin{center}
\begin{tikzpicture}[scale=.8,line width=1pt]
  \foreach \x in {1,3,5} {
    \draw[blue] (\x,0) ellipse (.8 and .4);
  };
  \foreach \x in {2,4} {
    \draw[line width=1.5pt] (\x-.7,0) -- (\x+.7,0);
  };
  \foreach \x in {.7,2.7,4.7} {
    \draw[green,fill] (\x,0) circle (.15);
  };
  \foreach \x in {1.3,3.3,5.3} {
    \draw[red,fill] (\x,0) circle (.15);
  };
  \draw[blue] (3,1.2) node (PS) {Physical spins $\cP$};
  \draw[blue,->,shorten >=1mm,overlay] (PS.west) -| (1,.4);
  \draw[blue,->,shorten >=1mm,overlay] (PS.east) -| (5,.4);
  \draw[blue,->,shorten >=1mm,overlay] (PS.south) -| (3,.4);
  \draw[green] (.6,0) ++(-1,1) node[text width=1.6cm,text centered]
    (BSL) {Boundary spin $\cV$};
  \draw[red] (5.3,0) ++(1,1) node[text width=1.6cm,text centered]
    (BSR) {Boundary spin $\cV^\ast$};
  \draw[green,->,shorten >=1.5mm] (BSL.south) |- (.7,0);
  \draw[red,->,shorten >=1.5mm] (BSR.south) |- (5.3,0);
\end{tikzpicture}
\end{center}
\caption{\label{fig:VBS}(Color online) The AKLT construction and the
  emergence of boundary spins. The ellipses denote the projection from
  $\cV\otimes\cV^\ast$ to $\cP$, the links between neighboring physical
  spins refer to singlet bonds.}
\end{figure}

In this article, we will solely be concerned with spin chains whose
spins transform in the adjoint representation $\theta$ of SU(N). This
representation has dimension $N^2-1$ and it is represented by the
Young tableau
\begin{align}
  \theta
  =\left.\tinyyng{2,1,1,1}\right\}{\scriptstyle N{-}1}\ \ .
\end{align}
Let us first discuss the most general form of translation invariant
nearest-neighbor Hamiltonians for this particular choice of spin.
Restricting our attention to two physical sites for a moment, the
Hilbert space $\theta\otimes\theta$ decomposes according to (sketch
for $N=5$)

\begin{equation}
\begin{split}
  \label{eq:Decomposition}
  \tinyyng{2,1,1,1}\otimes\tinyyng{2,1,1,1}
  &=\bullet\oplus2\;\tinyyng{2,1,1,1}
    \oplus\tinyyng{2,2,1}
    \oplus\tinyyng{3,1,1}
    \oplus\tinyyng{3,3,2,2}
    \oplus\tinyyng{4,2,2,2}\ \ .
\end{split}
\end{equation}

Just as indicated here, there are precisely seven contributions on the
right hand side provided $N$ satisfies the inequality $N\geq4$, see
Table~\ref{tb:RepsSUNadjoint} for the general result.\footnote{For
  $N=2$ the tensor product has only three parts while for $N=3$ there
  are six.} We also note the curious fact that the adjoint
representation appears with multiplicity two. One of the adjoints
resides in the symmetric part of the tensor product, the other one in
the anti-symmetric part. The occurrence of this non-trivial
multiplicity will play a crucial role later on, both technically and
physically. As will become clear below, these two adjoint
representations are associated with the completely symmetric tensor
$d^{abc}$ and the structure constants $f^{abc}$, respectively.

From the decomposition~\eqref{eq:Decomposition} we infer that there
are nine linearly independent SU(N) invariant operators which act on
the two-site Hilbert space. Out of these, seven are associated with
projectors $\bP_i$ onto the individual irreducible components
appearing in the decomposition \eqref{eq:Decomposition}. The labeling
of the projectors corresponds to that found in
Table~\ref{tab:AltSUN}. These operators all preserve the parity of the
tensor product, i.e.\ they have eigenvalue $+1$ under the action of
the permutation operator $\Pi$ which exchanges the two factors. Among
the $\bP_i$, the operators $\bP_S$ and $\bP_A$ will play a special
role since $\bP_S+\bP_A$ projects onto the two-dimensional space of
adjoints. In order to relate the associated spaces we also define a
new SU(N)-invariant operator $\bX$ which is characterized (up to a
sign) by the properties
\begin{align}
  \bX^2
  =\bP_S+\bP_A
  \quad\text{ and }\quad
  \{\bX,\Pi\}
  =0\;.
\end{align}
These equations state that $\bX$ is acting non-trivially only in the
subspace of the two adjoint representations where it exchanges the
symmetric and the anti-symmetric part. In Section~\ref{sc:InvTwoSites}
we construct explicit expressions for all the invariant operators
$\bP_i$ and $\bX$ and express them in terms of local spin operators.

The attentive reader may have realized that we only talked about eight
operators so far. The desired nine invariant operators are obtained if
$\bX$ is split into two nilpotent contributions according to
$\bX=\bX_{SA}+\bX_{AS}$ where the subscripts indicate which
irreducible representation space is mapped into which. The separate
parts $\bX_{SA}$ and $\bX_{AS}$, however, will not play a role in what
follows.

Our group theoretical analysis implies that the most general SU(N) and
translation invariant Hamiltonian with nearest-neighbor interactions
has nine parameters. Rescaling and shifting the energy leaving us with effectively seven
non-trivial coupling constants which could be parametrized by the
coordinates of the sphere~$S^7$. All our considerations will take
place in the phase diagram described by this sphere. In
particular, we ignore the possibility of staggered
couplings or long-range interactions.

\section{\label{sc:Haldane}Realization of Haldane phases}

\subsection{\label{sc:HaldaneGeneral}General considerations}

We shall now investigate whether the phase diagram under consideration
permits the construction of non-trivial Haldane phases. With a Haldane
phase we mean a gapped spin liquid phase of a 1D quantum spin system
which exhibits non-trivial symmetry protected topological order. As
was shown in Ref.~\onlinecite{Duivenvoorden:2012arXiv1206.2462D},
SU(N) spin systems can exhibit up to $N-1$ distinct Haldane phases. In
principle, all of them can be realized by using the ideas of the
original AKLT
construction.\cite{Affleck:PhysRevLett.59.799,Affleck:1987cy} Based on
this philosophy, examples of generalized AKLT Hamiltonians for SU(N)
have been worked out in
Refs.~\onlinecite{Affleck:1987cy,Rachel2010:PhysRevB.80.180420,Greiter:PhysRevB.75.184441,Duivenvoorden:2012arXiv1208.0697D,Nonne:2012arXiv1210.2072N,Morimoto:2014PhRvB..90w5111M,Quella:2015}.

The essential difference and technical complication of the present
case as compared to previous investigations in the literature is the
non-trivial multiplicity of the adjoint representation in the
decomposition~\eqref{eq:Decomposition}. As we will shortly review, the
AKLT recipe for the construction of a parent Hamiltonian requires one
to embed the tensor product of two auxiliary spins into the tensor
product of two physical spins. Since the former includes the adjoint
representation {\em once}, we encounter the mathematical problem of
identifying the proper one-dimensional subspace of this
two-dimensional space and of constructing the corresponding projection
operator (or rather its orthogonal complement). If we choose the wrong
embedding, the corresponding AKLT state fails to have the desired
properties, e.g.\ it might be degenerate.

AKLT states in the adjoint representation with the correct embedding
have first been discussed in
Ref.~\onlinecite{Duivenvoorden:2012arXiv1208.0697D} for the particular
case of an SU(3) spin chain. In what follows, we will generalize the
construction of Ref.~\onlinecite{Duivenvoorden:2012arXiv1208.0697D}
from SU(3) to SU(N), and we will also construct AKLT states based on
the auxiliary spin $\tinyyng{1,1}$ and its dual.

Before entering the technical details let us briefly outline the main
ingredients of a general AKLT construction for a symmetry group~$G$.

\begin{enumerate}
\item The physical spins transform in a representation $\cP$ which is
  assumed to be self-dual, $\cP^\ast=\cP$. Self-duality of the
  physical spin $\cP$ is required for the construction of a
  translation invariant AKLT state.
\item Each physical spin $\cP$ is thought to be made up from two
  auxiliary spins $\cV$ and $\cV^\ast$. In other words,
  $\cP\subset\cV\otimes\cV^\ast$ arises from a suitable projection on
  the tensor product of the two auxiliary spins.
\item For technical reasons we also demand that
  $\cV\otimes\cV^\ast\subset\cP\otimes\cP$. This should be thought of
  as a kind of non-degeneracy condition.
\end{enumerate}
These structures allow to write down an AKLT state
$|\text{AKLT}\rangle$ in matrix product form which is constructed from
the Clebsch-Gordan coefficients for the embedding
$\cP\subset\cV\otimes\cV^\ast$, see Figure~\ref{fig:VBS} for an
illustration. The state will be unique for an infinite system or if
periodic boundary conditions are imposed. In contrast, with open
boundary conditions on both sides it will be parametrized in terms of
two boundary spins $\cV$ and $\cV^\ast$ whose states are elements of
$\cV\otimes\cV^\ast$. The representation type of the boundary spins
will, moreover, determine the topological quantum number
characterizing the AKLT state (see
Ref.~\onlinecite{Duivenvoorden:2012arXiv1206.2462D} for details).

In addition to the state, the AKLT recipe also introduces a natural
$G$-symmetric and translation invariant Hamiltonian. For two physical
sites this Hamiltonian is given by
\begin{align}
  \label{eq:TwoSiteAKLT}
  h_{\text{AKLT}}
  \ =\ \idop-\bP_{\cV\otimes\cV^\ast}\;,
\end{align}
where $\bP_{\cV\otimes\cV^\ast}$ denotes the projection onto the
subspace $\cV\otimes\cV^\ast$ inside of $\cP\otimes\cP$. The
Hamiltonian $h_{\text{AKLT}}$ is then simply a projection onto its
orthogonal complement. The total Hamiltonian $H_{\text{AKLT}}$ is
given by summing $h_{\text{AKLT}}$ over all neighboring sites. By
construction, one has $H_{\text{AKLT}}\geq0$ and
$H_{\text{AKLT}}|\text{AKLT}\rangle=0$. Since this turns
$|\text{AKLT}\rangle$ into a ground state of $H_{\text{AKLT}}$ the
latter is usually called a parent Hamiltonian. Frequently,
$|\text{AKLT}\rangle$ is actually the unique ground state, at least
modulo the freedom of specifying the internal state of potentially
existing boundary spins.

Since it will be of great importance for our study, we would like to
emphasize a subtle point that arises if the representation $\cV$ is
not self-dual, i.e.\ if $\cV\neq\cV^\ast$. In this case, the AKLT
construction will generally lead to two different AKLT states
$|\text{AKLT}\rangle_1$ and $|\text{AKLT}\rangle_2$ which are
distinguished by the order of the factors in the tensor product
$\cV\otimes\cV^\ast$. While this tensor product is obviously invariant
under the exchange of $\cV$ and $\cV^\ast$, the 1D arrangement with
the presence of singlet bonds between neighboring physical sites
implies the presence of a natural orientation and a distinction in the
associated AKLT states. Mathematically speaking, the swap modifies the
embedding of $\cV\otimes\cV^\ast$ into $\cP\otimes\cP$. All our
statements can easily be illustrated with Figure~\ref{fig:VBS}, where
the exchange of $\cV$ and $\cV^\ast$ leads to the inversion of the
whole picture.  This also means that the state $|\text{AKLT}\rangle$
and the associated Hamiltonian $H_{\text{AKLT}}$ as defined from
Eq.~\eqref{eq:TwoSiteAKLT} will necessarily be {\em chiral} whenever
$\cV\neq\cV^\ast$.

Turning this argument around, this also implies the following
statement: If one has an inversion symmetric Hamiltonian which has one
of the AKLT states $|\text{AKLT}\rangle_1$ or $|\text{AKLT}\rangle_2$
as its ground state, then its image under inversion will also
automatically be a ground state. In other words: There is a
non-trivial ground state degeneracy and spontaneous inversion symmetry
breaking. This mechanism provides a general explanation for the
observations in the papers
\onlinecite{Affleck:1991NuPhB.366..467A,Greiter:PhysRevB.75.184441,Rachel:2010JPhCS.200b2049R,Morimoto:2014PhRvB..90w5111M}.

\subsection{Auxiliary spins in general anti-symmetric representations
  of SU(N)}

From now on we will assume the physical spin $\cP=\theta$ to transform
in the adjoint representation of SU(N). On the other hand, we will
explore the possibility to employ different types of auxiliary spins
$\cV$ for the AKLT construction, notably those of the form
\begin{align}
  \cV=\Sigma_l=\left.\tinyyng{1,1,1}\right\}{\scriptstyle l}
  \quad\text{ with }\quad
  \cV^\ast=\Sigma_{N-l}=\left.\tinyyng{1,1,1,1}\right\}{\scriptstyle N{-}l}\ \ .
\end{align}
The representation $\Sigma_l$ corresponds to the anti-symmetric part of
the $l$-fold tensor product of the fundamental representation with
itself. For this type of auxiliary spin, the dual representation is
given by $\Sigma_l^\ast=\Sigma_{N-l}$, i.e.\ it belongs to the same
family of representations. For symmetry reasons we can thus restrict
our attention to cases with $l\leq N/2$. As explained in
Section~\ref{sc:HaldaneGeneral}, the remaining cases can be obtained
by space inversion.

For the construction of the associated AKLT Hamiltonian we need to
know the tensor product decomposition of auxiliary spins which, in
this case, reads (for $1\leq l\leq N/2$)
\begin{align}
  \label{eq:TensorProductAux}
  \Sigma_l\otimes\Sigma_{N-l}
  =0\oplus\theta\oplus\bigoplus_{r=2}^l\Lambda_r
  =\bullet\oplus\theta\oplus\bigoplus_{r=2}^l\tinyyoung{\,1,\,\,,\,r,\,,\,}\;,
\end{align}
where the symbols $0$ and $\bullet$ both refer to the trivial
one-dimensional representation and the representations $\Lambda_r$ are
described by the Young tableaux
\begin{align}
  \Lambda_r
  ={\scriptstyle N{-}r}
   \left\{\tinyyoung{\,1,\,\,,\,r,\,,\,}\right.
   \begin{array}{ll}\!\!\!\Bigl\}\,{\scriptstyle r}\\[5mm]\end{array}
\end{align}
Formally, we can identify $\Lambda_0=0$ and $\Lambda_1=\theta$. The
representations we just encountered are summarized in
Table~\ref{tb:RepsSUNadjoint}, together with their most important
properties. We exclude the case $l=0$ since it violates condition~2,
i.e.\ it fails to feature the physical representation $\theta$ in the
tensor product \eqref{eq:TensorProductAux}.

We note that the tensor product
decomposition~\eqref{eq:TensorProductAux} has $l+1$ contributions. In
general it will thus not be possible to satisfy requirement~3, i.e.\
$\cV\otimes\cV^\ast\subset\cP\otimes\cP$, since the right hand side
only features up to seven representations. A closer inspection reveals
that condition~3 is satisfied if and only if $l=1$ or $l=2$. We will
thus restrict our attention to these two special cases in what follows
which will lead to Haldane phases with topological quantum number
$\pm[1]$ and $\pm[2]$ (if the image under space inversion is also
taken into account). The associated representation $\Sigma_l$ of
dimension $N$ (for $l=1$) or $\tfrac{1}{2}N(N-1)$ (for $l=2$) will be
referred to as the fundamental and the rank-2 anti-symmetric
representation, respectively. All other cases presumably suffer from
unwanted ground state degeneracies. However, a more detailed analysis
of this issue is beyond the scope of the present paper.

\begin{table}
\begin{center}
  \footnotesize
  \begin{tabular}{c|c|c|c|c}
    & $0$ & $\Sigma_l$, $\Sigma_l^\ast$ & $\theta$ & $\Lambda_l$ \\\hline&&&&\\[-3mm]
  Dimension & $1$ & $\frac{N!}{l!(N-l)!}$ & $N^2-1$ & $\frac{(N-2l+1)N!(N+1)!}{(l!(N+1-l)!)^2}$ \\[2mm]
  Casimir $\bQ$ & $0$ & $\frac{l}{N}(N+1)(N-l)$ & $2N$ & $2l(N+1-l)$ \\[2mm]
  Tableau & $\bullet$ & $l\left\{\tinyyng{1,1,1}\right.$ , $\left.\tinyyng{1,1,1,1}\right\}N{-}l$ &
  $\tinyyng{2,1,1,1,1,1}$ & $N{-}l\left\{\tinyyng{2,2,2,1}\right.\begin{array}{ll}\!\!\!\biggl\}\,l\\[3mm]\end{array}$
  \end{tabular}
  \caption{\label{tb:RepsSUNadjoint}Some representations of SU(N)
    and their data.}
\end{center}
\end{table}

\subsection{\label{sc:InvTwoSites}Invariant operators on two physical
  sites}

By construction, the two-site AKLT Hamiltonian~\eqref{eq:TwoSiteAKLT}
is an SU(N) invariant operator. We will express it in terms of the
seven projectors $\bP_i$ and the `permutation operator' $\bX$, see
Table~\ref{tab:AltSUN} for our labeling conventions. In terms of these
operators and choosing $\cV=\Sigma_1$ (the fundamental
representation), the most general candidate for the projection onto
the subspace $\cV\otimes\cV^\ast\subset\cP\otimes\cP$ is given by
\begin{equation}
  \label{eq:Projector1}
\begin{split}
  \bP
  &=\bP_{\bullet}+\cos^2(\theta)\,\bP_S+\sin^2(\theta)\,\bP_A
    +\tfrac{1}{2}\sin(2\theta)\,\bX\;.
\end{split}
\end{equation}
Choosing a convenient basis in the subspace of adjoints, the last four
terms can simply be written as the matrix\footnote{The sign is basis
  dependent therefore cannot be determined unambiguously.}
\begin{align}
  \label{eq:Embedding}
  \bP_{\text{ad}}
  =\mat\cos^2(\theta)&\tfrac{1}{2}\sin(2\theta)\\\tfrac{1}{2}\sin(2\theta)&\sin^2(\theta)\tam\;,
\end{align}
thereby making explicit its nature as a rank-1 projector. In other
words, the angle~$\theta$ parametrizes possible one-dimensional
subspaces in the two-dimensional space of adjoints. For the
construction of the AKLT Hamiltonian we need to identify the proper
value of the angle~$\theta$. This problem will be solved in
Section~\ref{sc:ParentHamiltonian} below.

The case where $\cV=\Sigma_2$ (the rank-2 anti-symmetric
representation) is very similar. Here one simply needs to add one more
projector $\bP_{S_2}$, resulting in
\begin{equation}
  \label{eq:Projector2}
\begin{split}
  \bP
  &=\bP_{\bullet}+\cos^2(\theta)\,\bP_S+\sin^2(\theta)\,\bP_A
    +\tfrac{1}{2}\sin(2\theta)\,\bX+\bP_{S_2}\;.
\end{split}
\end{equation}
Of course, for this case the correct angle $\theta$ will generally be
different than in the previous setup.

While giving the correct and complete answer for the Hamiltonian and
while even being suitable for an implementation on the computer (see
Appendix~\ref{sc:Projectors}), the invariant operators $\bP_i$ and
$\bX$ arose from rather abstract considerations and are thus not very
explicit. Another, physically more intuitive, basis of invariant
operators is provided by so-called Casimir operators which are
associated with invariant tensors.

The Hamiltonian expressed in terms of invariant operators $\bP_i$ and
$\bX$, is quite simple and suitable for an implementation on the
computer (see Appendix~\ref{sc:Projectors}). However it is physically
more intuitive to use a basis of the so-called Casimir operators which
are written in terms of the spins. In order to define the relevant
tensors we fix an arbitrary basis $S^a$ of anti-hermitean $su(N)$
generators, i.e.\ of traceless $N\times N$ matrices. These generators
satisfy the commutation relations
\begin{align}
  [S^a,S^b]={f^{ab}}_cS^c\;.
\end{align}
The first tensor we consider is the Killing form
\begin{align}
  \kappa^{ab}
  \ =\ \tr\bigl(S^aS^b\bigr)
\end{align}
and its inverse $\kappa_{ab}$. The tensors $\kappa^{ab}$ and
$\kappa_{ab}$ can be used to raise and lower indices. We also
introduce two distinct rank-3 tensors, namely
\begin{align}
  \label{eq:Tensors}
  f^{abc}
  &=\tr\bigl(S^a[S^b,S^c]\bigr)\quad\text{and}\\[2mm]
  d^{abc}
  &=\tr\bigl(S^a\{S^b,S^c\}\bigr)\;.
\end{align}
By construction $d^{abc}$ is completely symmetric while $f^{abc}$ is
completely anti-symmetric.

The two main examples of Casimir operators, and also the only ones
that will be needed in our paper, are the quadratic and the cubic
Casimirs
\begin{align}
  \bQ=\vec{S}^2
   =\kappa_{ab}S^aS^b
  \quad\text{ and }\quad
  \bC\ =\ d_{abc}S^aS^bS^c\;.
\end{align}
While $\bC$ is identically zero for SU(2), it is a non-trivial
invariant operator for all SU(N) with $N\geq3$.

Let us now focus our attention on the two-site Hilbert space
$\cP\otimes\cP$ for which the spin $\vec{S}=\vec{S}_1+\vec{S}_2$ is
the sum of the two individual spins. On the tensor product we define
the invariant operators
\begin{equation}
\begin{split}
  \label{eq:CSCA}
  \bC_S
  &\ =\ d_{abc}\bigl(S_1^aS_1^bS_2^c+S_1^aS_2^bS_2^c\bigr)\\
  \bC_A
  &\ =\ d_{abc}\bigl(S_1^aS_1^bS_2^c-S_1^aS_2^bS_2^c\bigr)\;.
\end{split}
\end{equation}
We note that the last operator is anti-symmetric under the exchange of
the two tensor factors. We also define
\begin{align}
  \bK\ =\ d_{abc}d_{def}S_1^aS_1^dS_1^eS_2^fS_2^bS_2^c\ \ .
\end{align}
One may easily convince oneself of the relations
\begin{equation}
\begin{split}
  \bQ&=\bQ_1+\bQ_2+2\vec{S}_1\cdot\vec{S}_2
      =4N+2\vec{S}_1\cdot\vec{S}_2\;,\\[2mm]
  \bC&=\bC_1+\bC_2+3\bC_S
    =3\bC_S\;.
\end{split}
\end{equation}
Here we used that $\bQ_1=\bQ_2=2N$ and $\bC_1=\bC_2=0$ for the adjoint
representation. The vanishing of $\bC_1$ and $\bC_2$ is simply a
consequence of the self-duality of the adjoint representation.

A complete basis in the nine-dimensional space of invariant operators
on the tensor product $\cP\otimes\cP$ is provided by
\begin{align}
  \label{eq:AllOps}
  \idop,\bQ,\bQ^2,\bQ^3,\bQ^4,\bC_S,\bC_A,\bK,\{\bK,\bC_A\}\;.
\end{align}
The eigenvalues of these operators can be calculated using a graphical
calculus (see Section~\ref{sc:Birdtracks}) and they are summarized in
Table~\ref{tab:AltSUN}. This table explicitly states how these
operators may be expressed in terms of the projectors $\bP_i$ and
$\bX$. Implicitly, it thus also contains information about all the
algebraic relations between these operators.

Let us briefly outline why the expressions in \eqref{eq:AllOps} indeed
provide a linearly complete independent set of invariant operators. An
inspection of Table~\ref{tab:AltSUN} for instance implies the relation
\begin{align}
  \bQ(\bQ-2N)(\bQ-4N)(\bQ-4N-4)(\bQ-4N+4)=0\;.
\end{align}
This explains the possibility to restrict one's attention to
polynomials in $\bQ$ of order less or equal to four. Other relations
allow for the simplification of mixed products such as $\bK\bQ$ as a sum
over a multiple of $\bK$ and a polynomial in $\bQ$. In particular, one
easily confirms that $\bK \bC_S=\bC_S\bC_A=0$.

From Table~\ref{tab:AltSUN} we can also read off the important
relation
\begin{equation}
\begin{split}
  \bK&=2N^2(N^2-4)\bP_{\bullet}-8N^2\bP_S\\[2mm]
   &\quad+2N^2(N-2)\bP_{S_1}-2N^2(N+2)\bP_{S_2}\;,
\end{split}
\end{equation}
which may be solved for the projector $\bP_S$, thereby giving rise to
\begin{align}
  \label{eq:PK}
  \bP_S
  =-\tfrac{1}{8N^2}\bK+\tfrac{N^2-4}{4}\,\bP_{\bullet}
   +\tfrac{N-2}{4}\,\bP_{S_1}
   -\tfrac{N+2}{4}\,\bP_{S_2}\;.
\end{align}

We emphasize that $\bQ$, $\bC_S$ and $\bK$ are all symmetric under the
exchange of $\vec{S}_1$ and $\vec{S}_2$. These operators hence
preserve the symmetric and the anti-symmetric part of the tensor
product $\cP\otimes\cP$. It is thus clear that we will have to use the
anti-symmetric operator $\bC_A$ for the construction of the
``permutation operator'' $\bX$. Their precise relation turns out to
be\footnote{Our methods only allow us to fix this relation up to a
  sign. As will be explained in Section~\ref{sc:Hamiltonians},
  changing this sign exchanges the roles played by the auxiliary spins
  $\cV$ and $\cV^\ast$. Since this is of limited physical significance
  we will ignore this issue in what follows.}
\begin{align}
  \label{eq:XCA}
  \bX
  &=\pm\tfrac{1}{2N\sqrt{N^2-4}}\,\bC_A\;.
\end{align}
To prove this relation we observe that both sides of the equation
anti-commute with the permutation operator $\Pi$, and compute the
square of each side.  More details may be found at the end of the
appendix.

\subsection{\label{sc:Birdtracks}Birdtracks}

The projectors~\eqref{eq:Projector1} and~\eqref{eq:Projector2} can
easily be constructed once one knows the eigenvalues of the invariant
operators $\vec{S}_1\cdot\vec{S}_2$, $\bC_A$ and $\bK$ (see
Table~\ref{tab:AltSUN}). To compute the eigenvalues of these invariant
operators, we use a diagrammatic method called {\it birdtracks}, which
was (re)discovered and refined by Cvitanovic.\cite{Cvitanovic} His
book contains the diagrammatic form of the projectors $\bP_i$ for the
various irreducible representations. We convert $\bK,\bC_A$ etc.\ into
diagrams, and find the answer for general values of $N$ after some
manipulation.

Throughout the paper we draw the fundamental and the adjoint
representation with straight and wavy lines, respectively. We will
therefore think of the $su(N)$ matrices $T^a$ in the fundamental
representation as a three-index object represented by the diagram
\begin{align}
(T^a)_{\alpha\beta}
=\ \includediagram{1.5cm}{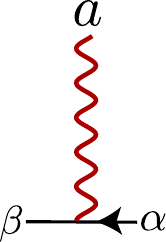}\ \ .
\end{align}
The fundamental and the anti-fundamental representation are
distinguished by the orientation of their arrow. Since the adjoint
representation is self-conjugate, there is no arrow on the wavy
lines. In the rest of this paper, we usually omit the indices -- after
all, this is the point of using diagrams in the first place.

As an illustration we wish to depict the defining property
$\tr(T^a)=0$ and the normalization convention $\tr(T^a
T^b)=\delta^{ab}$ in terms of diagrams. They simply read
\begin{align}
\label{eq:TraceNormalization}
\includediagram{1.5cm}{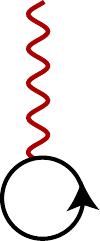}=0
\qquad\text{ and }\qquad
\includediagram{1.5cm}{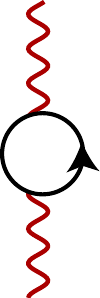}
\ =\ \includediagram{1.5cm}{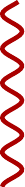}\ \ .
\end{align}
Similarly, the definition~\eqref{eq:Tensors} of the structure
constants $f^{abc}$ and the completely symmetric tensor $d^{abc}$
translates into the pictures
\begin{align}
\includediagram{1.5cm}{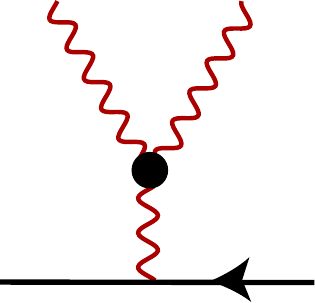}
&=\includediagram{1.5cm}{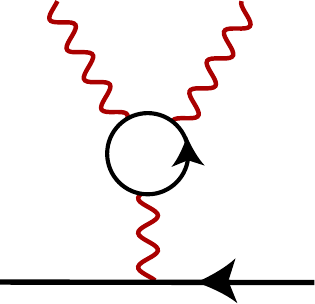}
-\includediagram{1.5cm}{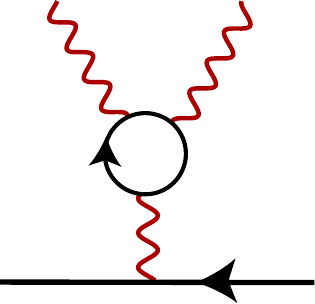} \label{eq:ftensor} \\[2mm]
\includediagram{1.5cm}{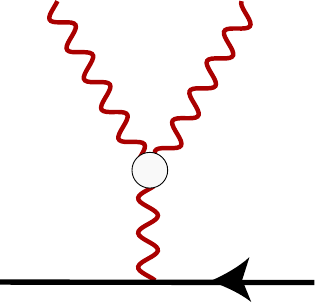}
&=\includediagram{1.5cm}{figs/appeq7.pdf}
+\includediagram{1.5cm}{figs/appeq8.pdf}\;,
\end{align}
Here we represented the tensors $f^{abc}$ and $d^{abc}$ by a filled
and an open circle, respectively.

Now imagine a complicated diagram composed of straight and wavy lines
(the former with arrows). These so-called birdtracks can be simplified
and evaluated by means of two elementary identities. The first of
these identities is
\begin{align}
\label{eq:FundId1}
\includediagram{1.5cm}{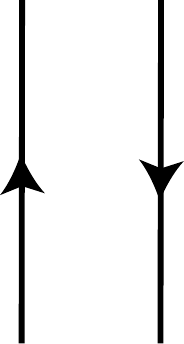}
=\frac{1}{N}\;\includediagram{1.5cm}{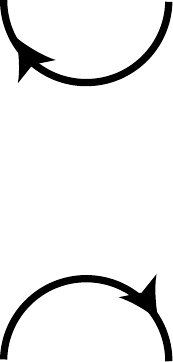}
\ +\ \includediagram{1.5cm}{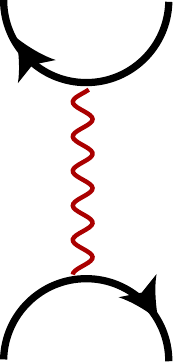}\;.
\end{align}
It simply expresses the familiar fact that
$\cV\otimes\cV^\ast=0\oplus\theta$ and decomposes the identity
operator on $\cV\otimes\cV^\ast$ into the corresponding projectors
onto the trivial (the trace part) and the adjoint representation.
From the left relation in Eq.~\eqref{eq:TraceNormalization} we see
that the prefactor $N$ in Eq.~\eqref{eq:FundId1} is required in order
to ensure the correct trace of the identity operator on $\cV$. This
trace $\tr_\cV(\idop)=N$ has a graphical representation as a solid
loop. The second of the fundamental identities
\begin{align}
\label{eq:FundId2}
\includediagram{1.5cm}{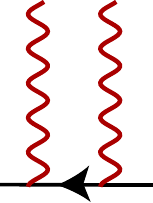}
-\includediagram{1.5cm}{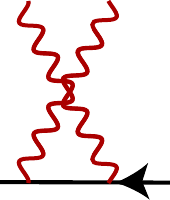}
=\includediagram{1.5cm}{figs/appeq9.pdf}
\end{align}
simply implements the Lie algebra commutation relations. Note that
there is no equivalent of this property for $d^{abc}$ (empty circle).

Let us illustrate the previous discussion with an example. A
diagrammatic equation that will be used in later sections is
\begin{align}
\label{eq:Traces}
\includediagram{1.0cm}{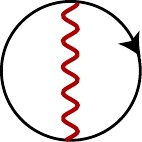}
=\tfrac{N^2-1}{N}\,\includediagram{1.0cm}{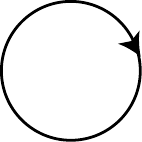}
=N^2-1
=\tr\left(\includediagram{1.0cm}{figs/appeq4.pdf}\right)\;.
\end{align}
It has two interpretations. The right hand side simply determines the
trace in the adjoint representation while on the left hand side we
calculate the trace of the quadratic Casimir operator in the
fundamental representation.

As pointed out in Section~\ref{sc:Setup}, the tensor product
$\cP\otimes\cP$ admits a nine-dimensional space of invariant operators
when $\cP$ is the adjoint representation and $N\geq4$. This space is
spanned by the invariant operators associated with the diagrams
\begin{align}
\label{eq:StandardBasis}
\includediagram{1.5cm}{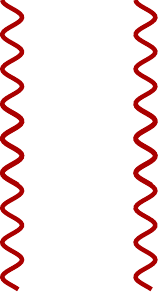}\,
\includediagram{1.5cm}{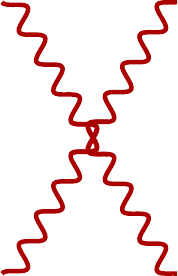}\,
\includediagram{1.5cm}{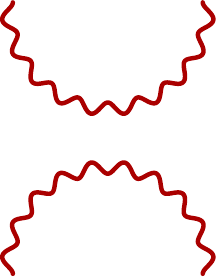}\,
\includediagram{1.5cm}{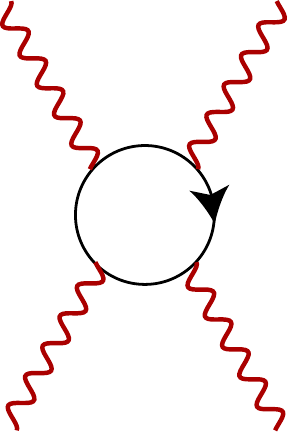}\,
\includediagram{1.5cm}{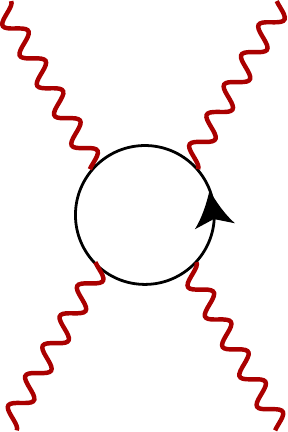}\\
\includediagram{1.5cm}{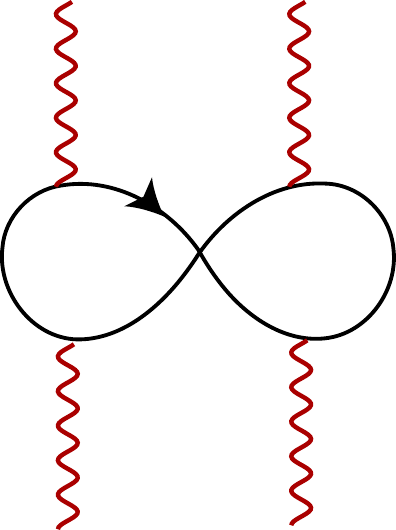}\,
\includediagram{1.5cm}{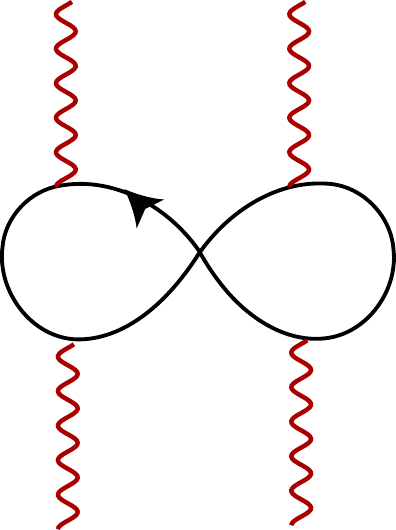}\,
\includediagram{1.5cm}{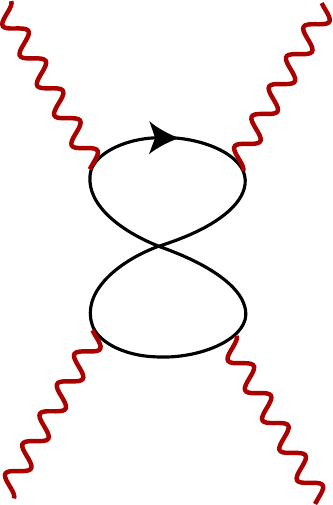}\,
\includediagram{1.5cm}{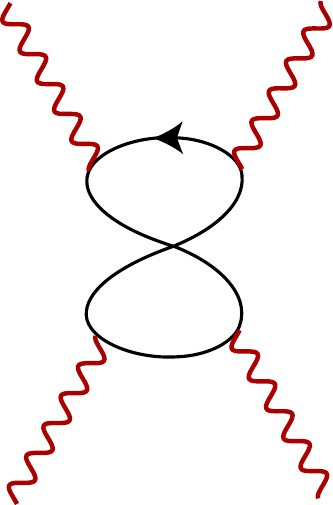}\;. \label{eq:diaglist}
\end{align}
On the top row we recognize the first three diagrams to be the
identity, the permutation and the (unnormalized) projection onto the
singlet. It can be shown that any birdtrack with four external wavy
lines and any combination of internal wavy and straight lines can be
reduced to a linear superposition of the previous diagrams. In
particular, there are expressions for the projectors $\bP_i$ and $\bX$
and we list them in Appendix~\ref{sc:Projectors} for later
convenience.

Since any SU(N) representation arises in a multiple tensor product of
the fundamental representation with itself, the graphical rules
considered so far are, in principle, sufficient to deal with arbitrary
physical representations. In order to describe the rank-2
anti-symmetric representation we only need to be able to project into
the anti-symmetric part of two fundamental
representations. Graphically, the (anti-)symmetrization will be
indicated as follows,
\begin{align}
\includediagram{.9cm}{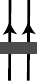}
=\includediagram{.9cm}{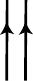}
-\includediagram{.9cm}{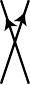},
\qquad
\includediagram{.9cm}{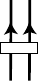}
=\includediagram{.9cm}{figs/appeq45.pdf}
+\includediagram{.9cm}{figs/appeq46.pdf}\;.
\end{align}
These two operations will project onto the representations
$\tinyyng{1,1}$ and $\tinyyng{2}$ respectively. Any Young tableau can
be represented by projectors in this way,\cite{Cvitanovic} but these
two are all we need in this paper.

{\em Inversion} and {\em Conjugation} (equivalent to time-reversal)
also have a diagrammatic interpretation. The former is reflection
about a vertical line and the latter corresponds to changing the
direction of every arrow. The two are {\em not} equivalent in
general. For example, the horizontal `figure-8' is invariant under one
but not the other.
\begin{align}
\begin{array}{ccc}
\includediagram{1.5cm}{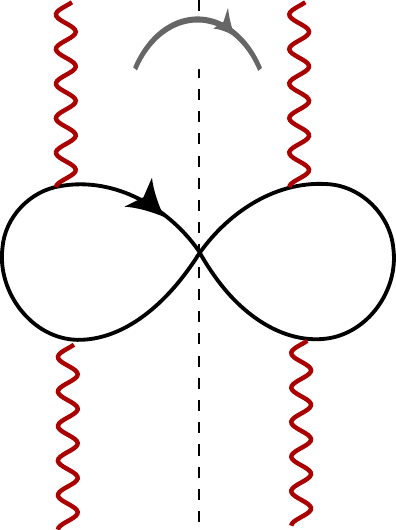}&
\xrightarrow{\mathbf{\tiny Inversion}}&
\includediagram{1.5cm}{figs/appeq26.pdf}\\
\includediagram{1.5cm}{figs/appeq26.pdf}&
\xrightarrow{\mathbf{\tiny Conjugation}}&
\includediagram{1.5cm}{figs/appeq27.pdf}
\end{array}
\end{align}

\subsection{Eigenvalues of invariant two-site operators}

Physical Hamiltonians are generally written in terms of spin
operators. The spin operator in the adjoint representation looks like
\begin{align}
  S_1^a
  \;=\;\includediagram{1.2cm}{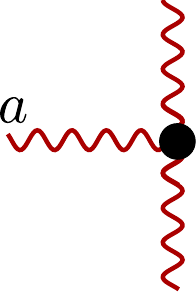}
  \;=\;-\includediagram{1.2cm}{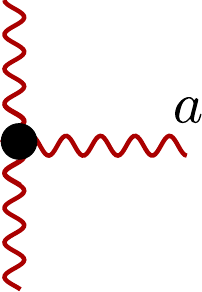} 
\end{align}
where the sign comes from the anti-symmetry in the structure constants
$f^{abc}$, see Eq.~\eqref{eq:ftensor}. The invariant operators
$\vec{S}_1\cdot\vec{S}_2$, $\bC_A$, $\bC_S$ and $\bK$ have been
introduced in Section~\ref{sc:InvTwoSites}. In diagrammatic form, some
of the relevant expressions read
\begin{widetext}
\begin{align}
-S_1\cdot S_2
&=\includediagram{1.2cm}{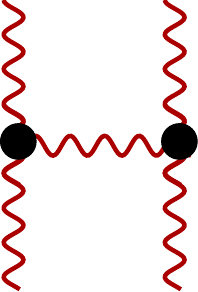}&
\frac{\bC_S+\bC_A}{2}=d_{abc}S_1^a S_1^b S_2^c
&=\includediagram{1.2cm}{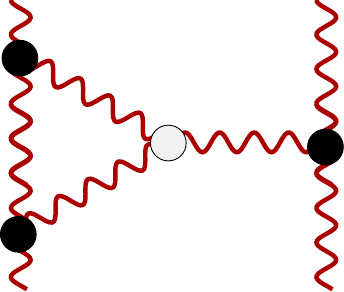}\\[2mm]
\frac{\bC_S-\bC_A}{2}=d_{abc}S_1^a S_2^b S_2^c
&=\,-\includediagram{1.2cm}{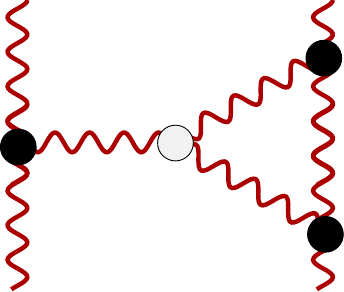}
&\bK=d_{abc}d_{efg}S_1^a S_1^f S_1^g S_2^e S_2^b S_2^c
&=\includediagram{1.2cm}{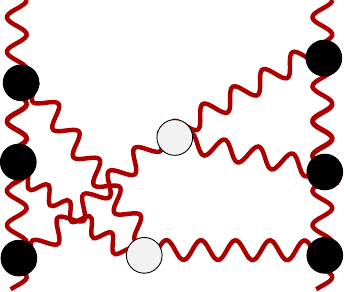}
\end{align}
These diagrams can be expressed in the standard
basis~\eqref{eq:StandardBasis} using the fundamental identities
\eqref{eq:FundId1} and \eqref{eq:FundId2}. After simplification, we
get sums over the terms listed in \eqref{eq:diaglist}. The final
expressions read
\begin{align}
\bC_S
&=2N\,\left\{
\includediagram{.9cm}{figs/appeq26.pdf}
-\includediagram{.9cm}{figs/appeq27.pdf}
\right\}\;,&
-S_1\cdot S_2
&=\left\{
\includediagram{.9cm}{figs/appeq23.pdf}
+\includediagram{.9cm}{figs/appeq23a.pdf}
\right\}
-\left\{
\includediagram{.9cm}{figs/appeq26.pdf}
+\includediagram{.9cm}{figs/appeq27.pdf}
\right\}\\[2mm]
\bC_A
&=2N\,\left\{
\includediagram{.9cm}{figs/appeq23a.pdf}
-\includediagram{.9cm}{figs/appeq23.pdf}
\right\}\;,&
\bK
&=-N^2\,\left\{
2\includediagram{.9cm}{figs/appeq15.pdf}
+2\includediagram{.9cm}{figs/appeq16.pdf}
-2\includediagram{.9cm}{figs/adjtrace.pdf}
-N\left(\includediagram{.9cm}{figs/appeq26.pdf}
+\includediagram{.9cm}{figs/appeq27.pdf}\right)
\right\}\;.
\end{align}
\end{widetext}

\section{\label{sc:Hamiltonians}AKLT Hamiltonians for the adjoint
  representation}

This section will be used to state the SU(N) AKLT Hamiltonians for
physical spins transforming in the adjoint representation. We will
distinguish two distinct cases with auxiliary spins in the fundamental
$\tinyyng{1}$ and in the rank-2 anti-symmetric representation
$\tinyyng{1,1}$, respectively. Two additional setups arise by
dualizing these representations which results in a simple change of
sign in the original AKLT Hamiltonians. We finally comment on the
special case SU(4) which is of particular interest since our
construction gives access to the complete set of three Haldane phases
in a single phase diagram.

\subsection{\label{sc:Fundamental}Haldane phases based on the
    fundamental representation}

\begin{table*}
	\begin{center}
		\footnotesize
		\begin{tabular}{c|ccccccc}
			$\tinyyng{2,1,1,1,1}\otimes\tinyyoung{11,2,\,,\,,r}$
			& $\bullet$
			& $\tinyyoung{\,1,1,2,3,4}$ 
			& $\tinyyoung{\,1,2,3,4,r}$
			& $\tinyyoung{\,11,2,3,4}$
			& $\tinyyoung{\,\,1,\,12,\,\,,\,\,,\,r}$\rule[-1.7em]{0pt}{1em}
			& $\tinyyoung{\,\,11,\,2,\,\,,\,\,,\,r}$
			& $\tinyyoung{\,1,12,3,4}$\\\hline\hline&&&&&&&\\[-1.2em]
			Symmetry & S & S & A & A & A & S & S\\
                        Projector & $\bP_{\bullet}$ & $\bP_S$ & $\bP_A$ & $\bP_{A_1}$ & $\bP_{A_2}$ & $\bP_{S_1}$ & $\bP_{S_2}$\\
			Dimension & $1$& $N^2{-}1$ & $N^2{-}1$ & $\frac{(N^2-1)(N^2-4)}{4}$ & $\frac{(N^2-1)(N^2-4)}{4}$ & $\frac{N^2(N+3)(N-1)}{4}$ & $\frac{N^2(N-3)(N+1)}{4}$  \\\hline&&&&&&&\\[-1em]
			$\bQ$ & $0$ & \multicolumn{2}{c}{$2N\smat1&0\\0&1\stam$} & $4N$ & $4N$ & $4(N+1)$ & $4(N-1)$\\[2mm]
			$\vec{S}_1\cdot\vec{S}_2$ & $-2N$ & \multicolumn{2}{c}{$-N\smat1&0\\0&1\stam$} & $0$ & $0$ & $2$ & $-2$\\\hline&&&&&&&\\[-1em]
			$\bC$ & $0$ &\multicolumn{2}{c}{$\smat0&0\\0&0\stam$} & $\pm12N$ & $\mp12N$ & $0$ & $0$ \\[2mm]
			$\bC_S$ & $0$ & \multicolumn{2}{c}{$\smat0&0\\0&0\stam$} & $\pm4N$ & $\mp4N$ & $0$ & $0$ \\[2mm]
			$\bC_A$ & $0$ &
			\multicolumn{2}{c}{$\pm 2N\sqrt{N^2-4}\smat0&1\\1&0\stam$} & $0$ & $0$
			& $0$ & $0$ \\[2mm]
			$\bK$ & $2N^2(N^2-4)$ & \multicolumn{2}{c}{$-8N^2\smat1&0\\0&0\stam$} & $0$ & $0$
			& $2N^2(N-2)$ & $-2N^2(N+2)$
		\end{tabular}
		\caption{\label{tab:AltSUN}The tensor product
                  $\theta\otimes\theta$ for SU(N) (rank $r=N-1$) for
                  $N\geq4$ and the eigenvalues of various invariant
                  operators on its individual irreducible
                  constituents. The row ``Symmetry'' indicates whether
                  the representation lies in the symmetric (S) or in
                  the anti-symmetric (A) part of the tensor
                  product. The correct signs in the expressions
                  involving a $\pm$ cannot be determined using our
                  methods but they have limited physical
                  significance.}
	\end{center}
\end{table*}

As discussed in Section~\ref{sc:HaldaneGeneral}, the two-site
interaction featuring the AKLT Hamiltonian is the projection onto the
orthogonal complement of $\cV\otimes\cV^\ast$ in $\cP\otimes\cP$. For
auxiliary spins in the fundamental representation (i.e.\
$\cV=\tinyyng{1}$) the relevant tensor product decomposition reads
\begin{align}
  \cV\otimes\cV^\ast
  \ =\ \bullet\oplus\theta\;.
\end{align}
This needs to be compared with the tensor product decomposition of
$\cP\otimes\cP$ which can be found in Table~\ref{tab:AltSUN}.

Our goal is to express the projector~\eqref{eq:Projector1} in terms of
spin operators and to determine the proper value of the angle
$\theta$. The latter problem is conveniently solved by means of the
birdtrack calculus which may be used to determine the correct
embedding of $\cV\otimes\cV^\ast$ into the physical two-site Hilbert
space $\cP\otimes\cP$. A detailed analysis (to be found in
Section~\ref{sc:ParentHamiltonian}) fixes the corresponding projection
matrix~\eqref{eq:Embedding} to be
\begin{align}
  \bP_{\text{ad}}
  =\frac{1}{2N^2-4}
   \mat N^2-4&\pm N\sqrt{N^2-4}\\\pm N\sqrt{N^2-4}&N^2\tam\;.
\end{align}
By resolving the matrix structure, this expression can be written
explicitly as a combination of the projectors $\bP_S$, $\bP_A$ and the
``permutation operator'' $\bX$. We then immediately obtain the
two-site Hamiltonian
\begin{align}
  \label{eq:AKLT1-1}
  h&=\idop-\Bigl(\bP_{\bullet}+\tfrac{N^2-4}{2(N^2-2)}\,\bP_S
     +\tfrac{N^2}{2(N^2-2)}\,\bP_A
     +\tfrac{N\sqrt{N^2-4}}{2(N^2-2)}\,\bX\Bigr)\;.
\end{align}

Finding an expression for this Hamiltonian which purely involves the
operators $\vec{S}_1\cdot\vec{S}_2$, $\bC_A$ and $\bK$ is a
straightforward but lengthy exercise. Instead of giving details of the
calculation we restrict ourselves to pointing out the general
procedure. We begin with writing the identity operator $\idop$ as a
sum over projectors and by splitting the contributions for $\bP_S$ and
$\bP_A$ into those of $\bP_S$ and $\bP_S+\bP_A$. This is advantageous
since $\bP_S+\bP_A$ has a simple expression in terms of
$\bQ=4N+2\vec{S}_1\cdot\vec{S}_2$ (see Appendix~\ref{sc:Projectors})
while $\bP_S$ can be rewritten in terms of $\bK$ using
Eq.~\eqref{eq:PK}. After also replacing $\bX$ by $\bC_A$ using
Eq.~\eqref{eq:XCA} one then simply needs to expand the polynomials in
$\vec{S}_1\cdot\vec{S}_2$ in order to arrive at
\begin{equation}
\begin{split}
  \label{eq:AKLT1-2}
  h&=\idop-\tfrac{1}{4N^2(N^2-2)}\,\bK
     -\tfrac{1}{4(N^2-2)}\,\bC_A
     +\alpha_1\,\vec{S}_1\cdot\vec{S}_2\\[2mm]
   &\quad+\alpha_2\,(\vec{S}_1\cdot\vec{S}_2)^2
     +\alpha_3\,(\vec{S}_1\cdot\vec{S}_2)^3
     +\alpha_4\,(\vec{S}_1\cdot\vec{S}_2)^4
\end{split}
\end{equation}
  with coefficients
\begin{align}
  \alpha_1
  &=\tfrac{N(N^2-3)}{4(N^2-1)(N^2-2)}&
  \alpha_2
  &=-\tfrac{N^2+1}{4(N^2-1)(N^2-2)}\\[2mm]
  \alpha_3
  &=\tfrac{N}{8(N^2-1)(N^2-2)}&
  \alpha_4
  &=\tfrac{1}{8(N^2-1)(N^2-2)}\;.
\end{align}
The AKLT Hamiltonian for the configuration with swapped auxiliary
spaces, $\cV\leftrightarrow\cV^\ast$, is identical to
\eqref{eq:AKLT1-2} except for a change of sign in front of
$\bC_A$. This change of sign indeed just reflects the effect of
inverting the order of the sites along the chain since $\bC_A$ is
anti-symmetric under the exchange of $\vec{S}_1$ and $\vec{S}_2$.

\subsection{\label{sc:Antisymmetric}Haldane phases based on the
  anti-symmetric representation}

The analysis for the case of auxiliary spins transforming in the
anti-symmetric rank-2 tensor completely parallels the previous
calculation. Superficially, the only difference is the presence of an
additional projector $\bP_{S_2}$ in
Eq.~\eqref{eq:Projector1}. However, apart from the need to incorporate
$\bP_{S_2}$, the adjoint representation in $\cV\otimes\cV^\ast$ is
also embedded differently into $\cP\otimes\cP$. A detailed analysis
(see again Section~\ref{sc:ParentHamiltonian}) results in the
projector
\begin{align}
  \bP_{\text{ad}}
  &=\xi\mat  (N+2)(N-4)^2&\pm N(N-4)\sqrt{N^2-4} \\ \pm N(N-4)\sqrt{N^2-4} & N^2(N-2) \tam
\end{align}
with $\xi=1/\bigl[2(N^2(N-4)+16)\bigr]$. The associated two-site
Hamiltonian is given by the expression
\begin{align}
  \label{eq:AKLT2-1}
  h&=\idop-\Bigl(\bP_{\bullet}+\tfrac{(N+2)(N-4)^2}{2(16+N^2(N-4))}\,\bP_S\\[2mm]
   &\qquad+\tfrac{N^2(N-2)}{2(16+N^2(N-4))}\,\bP_A
     +\tfrac{N(N-4)\sqrt{N^2-4}}{2(16+N^2(N-4))}\,\bX
     +\bP_{S_2}\Bigr)\nonumber
\end{align}
As before, the Hamiltonian can be rewritten in terms of spin
operators. After some lengthy calculations this leads to the final
form
\begin{align}
  \label{eq:AKLT2-2}
  h&\ =\ \idop
         +\beta_1\,\bK
         +\beta_2\,\bC_A
         +\alpha_1\,\vec{S}_1\cdot\vec{S}_2\\[2mm]
   &\qquad+\alpha_2\,(\vec{S}_1\cdot\vec{S}_2)^2
         +\alpha_3\,(\vec{S}_1\cdot\vec{S}_2)^3
         +\alpha_4\,(\vec{S}_1\cdot\vec{S}_2)^4\nonumber
\end{align}
with coefficients
\begin{align}
  \beta_1
  &=-\tfrac{N^2-8}{4N^2(16+N^2(N-4))}&
  \beta_2
  &=-\tfrac{N-4}{4(16+N^2(N-4))}\\[2mm]
  \alpha_1
  &=\tfrac{N(2N^3+N^2-2N-12)}{4(N+1)(16+N^2(N-4))}&
  \alpha_2
  &=-\tfrac{N^4-N^3-4N^2-8N+16}{8(N+1)(16+N^2(N-4))}\nonumber\\[2mm]
  \alpha_3
  &-\tfrac{3N^3+10N^2-20N-16}{16(N+1)(16+N^2(N-4))}&
  \alpha_4
  &=-\tfrac{N^2+4N-8}{16(N+1)(16+N^2(N-4))}\;.\nonumber
\end{align}
Just as in the previous case, the exchange of auxiliary spaces,
$\cV\leftrightarrow\cV^\ast$, leads to the opposite sign in the
contribution involving $\bC_A$.

\subsection{\label{sc:Cases}Discussion of special cases}

In this section we will discuss some peculiarities arising in the
special cases SU(3) and SU(4). For SU(3) it will be demonstrated that
we recover the Hamiltonian of
Ref.~\onlinecite{Duivenvoorden:2012arXiv1208.0697D}. We will also
explicitly analyze the case SU(4) which is the only instance where the
system with auxiliary spins in the rank-2 anti-symmetric representation
is invariant under space inversion.

\subsubsection{\label{sc:SU3}The case of SU(3)}

For SU(3), the tensor product of the adjoint with itself reads
\begin{align}
  \tinyyng{2,1}\otimes\tinyyng{2,1}
  \ =\ \bullet\oplus2\,\tinyyng{2,1}\oplus\tinyyng{4,2}
       \oplus\tinyyng{3}\oplus\tinyyng{3,3}\;.
\end{align}
This decomposition is slightly different than the generic one for
$N\geq4$ stated in Eq.~\eqref{eq:Decomposition} since one of the seven
representations, the representation $\cP_{S_2}$, is missing here. For
the remaining six projectors and for the AKLT Hamiltonians we may
nevertheless use the expressions that we derived for general values
of~$N$. In this case one may actually convince oneself that the two
Hamiltonians~\eqref{eq:AKLT1-1} and~\eqref{eq:AKLT2-1} reduce to the
same expression up to a sign in front of the operator $\bX$ (or
$\bC_A$), just as it should be.

A straightforward calculation yields the concrete form
\begin{equation}
  \label{eq:AKLTSU3}
\begin{split}
  h&=\idop-\Bigl(\bP_{\bullet}+\tfrac{5}{14}\bP_S+\tfrac{9}{14}\bP_A\pm\tfrac{3\sqrt{5}}{14}\bX\Bigr)\\[2mm]
   &=\idop-\tfrac{1}{252}\bK+\tfrac{9}{56}\,\vec{S}_1\cdot\vec{S}_2
     -\tfrac{5}{112}\,\bigl(\vec{S}_1\cdot\vec{S}_2\bigr)^2\\[2mm]
   &\qquad-\tfrac{1}{112}\,\bigl(\vec{S}_1\cdot\vec{S}_2\bigr)^3
     \mp\tfrac{1}{28}\,\bC_A
\end{split}
\end{equation}
for the two-site AKLT Hamiltonian. In contrast to general values of
$N$, this Hamiltonian only features spin-spin couplings up to third
order in $\vec{S}_1\cdot\vec{S}_2$. This is due to an additional
fourth order relation
\begin{align}
  \vec{S}_1\cdot\vec{S}_2\,(\vec{S}_1\cdot\vec{S}_2-2)(\vec{S}_1\cdot\vec{S}_2+3)(\vec{S}_1\cdot\vec{S}_2+6)=0
\end{align}
which arises from the absence of the seventh representation in the
tensor product (see Table~\ref{tab:AltSUN}). Needless to say, our
Hamiltonian~\eqref{eq:AKLTSU3} perfectly matches the result obtained
earlier in Ref.~\onlinecite{Duivenvoorden:2012arXiv1208.0697D}. In
order to compare with the results of
Ref.~\onlinecite{Duivenvoorden:2012arXiv1208.0697D} one needs the
identifications $\bC_A=8C_a$ and $\bK=16C^{(2)}$ which follow from the
respective tables for the eigenvalues.

\subsubsection{\label{sc:SU4}The case of SU(4)}

For SU(4), the physical two-site Hilbert space decomposes as
\begin{align}
  \tinyyng{2,1,1}\otimes\tinyyng{2,1,1}
  =\bullet\oplus2\,\tinyyng{2,1,1}\oplus\tinyyng{2,2}
        \oplus\tinyyng{4,2,2}
        \oplus\tinyyng{3,1}
        \oplus\tinyyng{3,3,2}\;.
\end{align}
On the level of the auxiliary representations one can only reach
\begin{align}
  \tinyyng{1}\otimes\tinyyng{1,1,1}
  \ =\ \bullet\oplus\tinyyng{2,1,1}
  \quad\text{ or }\quad
  \tinyyng{1,1}\otimes\tinyyng{1,1}
  \ =\ \bullet\oplus\tinyyng{2,1,1}
       \oplus\tinyyng{2,2}\;.
\end{align}
We note that the second option gives rise to an inversion symmetric
AKLT state. In both cases, the contributions are contained in the
two-site Hilbert space $\cP\otimes\cP$.

It remains to simplify the general expressions \eqref{eq:AKLT1-2} and
\eqref{eq:AKLT2-2} to the cases at hand. The simplified two-site
Hamiltonians read
\begin{equation}
\begin{split}
  \label{eq:AKLTSU4-1}
  h&=\idop-\frac{\bC_A}{56}-\frac{\bK}{896}
     +\tfrac{13}{210}\,\vec{S}_1\cdot\vec{S}_2
     -\tfrac{17}{840}(\vec{S}_1\cdot\vec{S}_2)^2\\[2mm]
   &\qquad+\tfrac{1}{420}\,(\vec{S}_1\cdot\vec{S}_2)^3
     +\tfrac{1}{1680}\,(\vec{S}_1\cdot\vec{S}_2)^4
\end{split}
\end{equation}
  for an auxiliary spin in the fundamental representation and
\begin{equation}
  \label{eq:AKLTSU4-2}
\begin{split}
  h&=\idop-\tfrac{1}{128}\,\bK +\tfrac{31}{20}\,\vec{S}_1\cdot\vec{S}_2
     -\tfrac{7}{40}\,(\vec{S}_1\cdot\vec{S}_2)^2\\[2mm]
   &\qquad-\tfrac{1}{5}\,(\vec{S}_1\cdot\vec{S}_2)^3
     -\tfrac{3}{160}\,(\vec{S}_1\cdot\vec{S}_2)^4
\end{split}
\end{equation}
for an auxiliary spin in the anti-symmetric representation.

We recognize that the second Hamiltonian does not feature the term
$\bC_A$ which is anti-symmetric under the exchange of the two spins
$\vec{S}_1$ and $\vec{S}_2$. As a consequence, the resulting AKLT
Hamiltonian is inversion symmetric. This property is related to the
self-duality of the representation $\tinyyng{1,1}$ which holds if and
only if $N=4$. We note that the Hamiltonians~\eqref{eq:AKLTSU4-1}
and~\eqref{eq:AKLTSU4-2} realize the full set of three Haldane phases
for SU(4) if one takes into account the possibility to reverse the
sign in front of $\bC_A$ in Eq.~\eqref{eq:AKLTSU4-1} which exchanges
the roles of $\cV$ and $\cV^\ast$.

\section{\label{sc:Correlations}Correlation lengths}

One of the characteristics of a physical system is the existence or
absence of a gap. We will now show that both models under
investigation are gapped by proving the existence of exponentially
decaying correlations. This will be done by diagonalizing the
associated transfer matrix and finding a gap between the largest (by
absolute value) two eigenvalues.

\subsection{Fundamental representation}

The transfer matrix~$E$ associated with our first AKLT state has the
following simple graphical representation
\begin{align}
  E=\includediagram{1.5cm}{figs/appeq2.pdf}\;.
\end{align}
The matrix~$E$ is a tensor with four indices and can, in principle, be
interpreted in a variety of ways. In the context of calculating the
norm of a state or correlation functions it is, however, convenient to
read this diagram from {\it left to right}, i.e.\ to interpret $E$ as
an operator on the space $\cV\otimes\cV^\ast$.

In the case of $\cV=\tinyyng{1}$ being the fundamental representation,
this tensor product decomposes as $\cV\otimes\cV^\ast=0\oplus\theta$
and hence the transfer matrix possesses a decomposition in terms of
projectors as $E=c_1\bP_1+c_2\bP_2$. The numbers $c_1$ and $c_2$ are
the eigenvalues of the transfer matrix which determine the correlation
length. In terms of our graphical calculus, this equation simply
translates into
\begin{align}
\includediagram{1.5cm}{figs/appeq2.pdf}
=c_1\,\frac{1}{N}\includediagram{1.5cm}{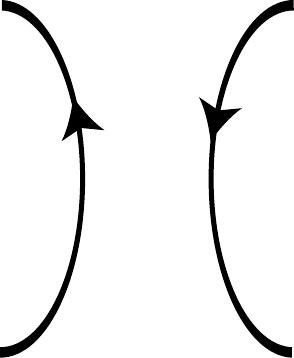}
 -c_2\includediagram{1.5cm}{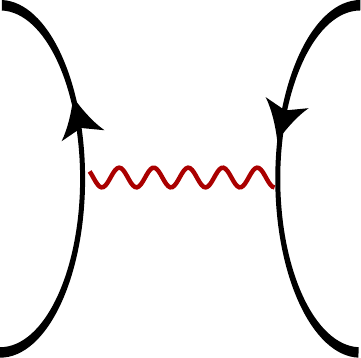}\;,
\end{align}
where the (normalized) diagrams correspond to the projectors $\bP_0$
and $\bP_\theta$ (compare with Eq.~\eqref{eq:FundId1}). The eigenvalue
$c_1$ can be determined by sandwiching both sides of the equation
between two right/left cups. According to
Eq.~\eqref{eq:TraceNormalization} this renders the rightmost diagram
trivial while Eq.~\eqref{eq:Traces} implies the value
$c_1=N-1/N$. Similarly, one finds a relation between $c_1$ and $c_2$
by sandwiching the original equation with two cups from the
top/bottom. We refer to these two operations as the horizontal and
vertical trace, respectively. Eventually we find the eigenvalues
$c_1=N-1/N$ and $c_2=-1/N$ with multiplicities $1$ and $N^2-1$
respectively. We note the strict inequality $|c_1|>|c_2|$ which proves
the desired spectral gap. From this simple and intuitive calculation
we can infer the correlation length as
\begin{align}
  \xi_{\tinyyng{1}}
  =1/\ln(|c_1/c_2|)=1/\ln(N^2-1)\;.
\end{align}
This result confirms earlier computations by various
groups.\cite{Katsura:2008JPhA...41m5304K,Rachel:2010JPhCS.200b2049R,Orus:2011PhRvB..83t1101O,Morimoto:2014PhRvB..90w5111M}

\subsection{Anti-symmetric representation}

While a neat exercise for the fundamental representation, the full
power of the graphical calculus becomes visible when considering more
complicated auxiliary spins such as the anti-symmetric representation
$\cV=\tinyyng{1,1}$. In this case, the tensor product
$\cV\otimes\cV^\ast=0\oplus\theta\oplus\cP_{S_2}$ decomposes into
three irreducible representations and the associated transfer matrix
reads $E=c_1\bP_1+c_2\bP_2+c_3\bP_3$. When interpreted with the help
of birdtracks, this equation reads
\begin{equation}
\begin{split}
  \includediagram{1.2cm}{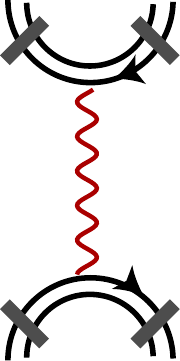}
  &=c_1\, \frac{2}{N(N-1)}\includediagram{1.2cm}{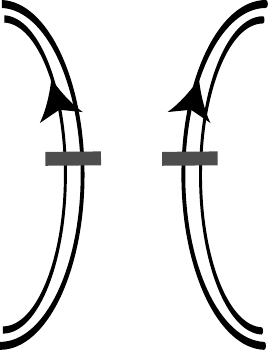}
    +c_2\, \frac{1}{N-2}\includediagram{1.2cm}{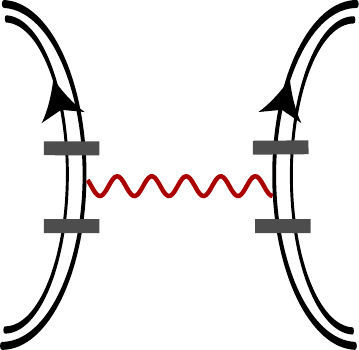}\\[2mm]
  &\qquad+c_3\, \alpha(N)\includediagram{1.2cm}{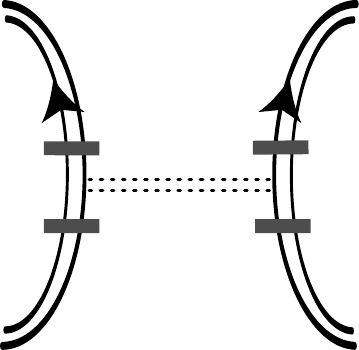}\;.
\end{split}
\end{equation}
Here, we introduced the dotted double line as a new notation for the
representation $\cP_{S_2}$ and $\alpha(N)$ is a normalization constant
whose precise value will not be of any concern. Again we need to
compute the eigenvalues $c_i$. Using the fact that the (horizontal)
trace is zero, we may do so without needing the exact form of the last
vertex or its normalization constant $\alpha(n)$. The eigenvalues are
\begin{align}
  c_1=\tfrac{2(N+1)}{N},\ 
  c_2=\tfrac{N^2-2N-4}{N(N-2)}
  \ \text{ and }\ 
  c_3=\tfrac{4}{N(N-2)}\;.
\end{align}
We can now extract the correlation length which is the ratio of the
next highest eigenvalue to the highest. We find
\begin{align}
  \xi_{\tinyyng{1,1}}
  =1/\ln\Bigl[\tfrac{N^2-2N-4}{2(N+1)(N-2)}\Bigr]\;.
\end{align}
We note that this correlation length (in contrast to
$\xi_{\tinyyng{1}}$) has a non-trivial finite value in the limit
$N\to\infty$.

\section{\label{sc:PhaseDiagram}Phase diagram}

The two AKLT Hamiltonians we discussed in
Sections~\ref{sc:Fundamental} and~\ref{sc:Antisymmetric} arise from
fine-tuning several coupling constants such as to clearly exhibit the
topological properties of the respective Haldane phases. It is a
natural question which other Haldane phases can possibly be realized
in the full space of SU(N) invariant Hamiltonians and where the phase
boundaries are located. In particular, it appears natural that the
topological phase transitions are described in terms of a critical
conformal field theory (CFT).

\subsection{Critical point with enhanced symmetry}

While a general discussion of phase transitions is beyond the scope of
the present paper we wish to point out that the phase diagram features
a critical point with enhanced $\text{SU($\cN$)}$ symmetry where we
introduced the abbreviation $\cN=N^2-1$. This point is associated with
a two-site Hamiltonian which degenerates into the permutation operator
or, equivalently, to the projection onto the symmetric part of the
two-site Hilbert space. This Hamiltonian can effectively be thought of
as an integrable Heisenberg Hamiltonian for the fundamental
representation of $\text{SU($\cN$)}$ whose dimension coincides with
the dimension of the adjoint representation of
SU(N).\cite{Sutherland:1975PhRvB..12.3795S} If $\Pi$ denotes the
permutation operator on two physical sites and $\vec{T}$ the spin
operator of $\text{SU($\cN$)}$, then the corresponding two-site
Hamiltonian can be written as
\begin{align}
  \label{eq:HamiltonianPermutation}
  h=\tfrac{1}{2}\bigl[\idop+\Pi\bigr]
   =\tfrac{1}{2}\Bigl[\vec{T}_1\cdot\vec{T}_2+\tfrac{N+1}{N}\,\idop\Bigr]\;.
\end{align}
As is well known, the associated universality class is described by a
$\text{SU($\cN$)}$ WZW model at level $k=1$. In terms of SU(N)
projectors, the Hamiltonian \eqref{eq:HamiltonianPermutation} has the
form $h=\bP_{\bullet}+\bP_S+\bP_{S_1}+\bP_{S_2}$ which can easily be converted into
the simple expression
\begin{align}
  h=-\tfrac{1}{8N^2}\,\bK
    +\tfrac{N}{8}\,\vec{S}_1\cdot\vec{S}_2
    +\tfrac{1}{8}\,(\vec{S}_1\cdot\vec{S}_2)^2
\end{align}
using the spin operators $\vec{S}$ of SU(N).

\subsection{Other integrable points}

It is conceivable that there are other Bethe ansatz solvable
Hamiltonians in the phase diagram of our model. A systematic
discussion of integrable SU(N) Hamiltonians based on completely
symmetric representations $\tinyyoung{1\,\,m}$ with $m$ boxes has been
provided by Andrei and Johannesson
\onlinecite{Andrei:1984PhLA..104..370A,Johannesson:1986NuPhB.270..235J,Alcaraz:1989JPhA...22L.865A}.
From the thermodynamic Bethe ansatz it could be inferred that these
systems are critical and in the universality class of the SU(N) WZW
model at level $k=m$. In principle, the approach of fusing the
R-matrix of the fundamental representation to obtain integrable models
for higher spins can be generalized to the adjoint
representation. However, since the procedure is quite technical we
will leave this analysis to future work. We believe that the corresponding
theory flows to the critical SU(N) WZW model at level $k=N$.

\subsection{Multicritical points from symmetry protection}

The universality classes of critical SU(N) quantum spin systems in one
dimension are provided by SU(N) WZW models.\cite{Witten:1983ar} These
WZW models are parametrized by a number $k=1,2,\ldots$, commonly known
as the level. The level can be thought of as a measure for the degree
of multicriticality. In general, WZW models with $k\geq2$ are unstable
and flow to WZW models with smaller values of
$k$.\cite{Affleck:1985wb,Affleck:1988NuPhB.305..582A} This is due to
the presence of non-trivial relevant operators which are compatible
with the SU(N) symmetry and which may trigger renormalization group
flows.

A recent study by Furuya and Oshikawa has shown that the presence of
additional discrete symmetries implies strong selection rules on
renormalization group flows in $\text{SU(2)}_k$ WZW
theories.\cite{Furuya:2015arXiv150307292F} Specifically, these WZW
models have a $\Integer_2$ symmetry which may be gauged for even
values of the level (turning the SU(2) WZW model into an SO(3) WZW
model) but which is anomalous for all odd values of $k$. As a
consequence, the level is preserved modulo~$2$ if a renormalization
group flow is triggered by relevant operators respecting this
$\Integer_2$-symmetry.\cite{Furuya:2015arXiv150307292F} This means
that there are in fact two stable fixed-points, namely $k=1$ and
$k=2$. Here, the stability of the $\text{SU(2)}_2$ WZW model requires
an additional $\Integer_2$ symmetry to be preserved.

The arguments of Ref.~\onlinecite{Furuya:2015arXiv150307292F} can
easily be generalized to other symmetry groups using the general
classification of orbifold WZW
theories.\cite{Felder:1988CMaPh.117..127F} For the group of interest
for us, namely SU(N), the corresponding analysis has been carried out
in Ref.~\onlinecite{Lecheminant:2015arXiv150901680L}. In that case,
the discrete symmetry which needs to be gauged is the group
$\Integer_N$, the center of SU(N). This effectively turns the group
SU(N) into the group $\text{PSU(N)}=\text{SU(N)}/\Integer_N$.

Let us now try a change of perspective. So far, the discussion in the
literature was solely concerned with the stability of gapless phases
and with selection rules on renormalization group flows between
conformal field theories. Here, we would like to point out that the
same arguments also have strong implications for the nature of
topological phase transitions. More precisely, we will argue that the
phase transitions between symmetry protected topological phases of
SU(N) lattice models may correspond to multicritical points, i.e.\ to
higher level SU(N) WZW models, under specific circumstances.

To explain the basic philosophy underlying our claim we first look at
the familiar example of the spin-1 Haldane phase for SU(2). Since the
spins transform in the spin-1 representation, the subgroup
$\Integer_2\subset\text{SU(2)}$ is acting trivially, turning the
actual symmetry into $\text{SO(3)}=\text{SU(2)}/\Integer_2$. In other
words, the model has an additional $\Integer_2$ symmetry (or rather
redundancy) which needs to be present everywhere in the phase diagram
as long as no extra degrees of freedom with half-integer spin are
added. This $\Integer_2$-invariance then of course should also exist
in the CFT describing the topological phase transition from the
Haldane phase to the dimerized phase. And indeed, this transition is
coinciding with the integrable Babujian-Takhtajan model which is
well-known to be described by an $\text{SU(2)}_2$ WZW model (see
Ref.~\onlinecite{Affleck:1985wb} and references therein). In the
previous argument we silently skipped over a subtle point. In fact, as
explained in Ref.~\onlinecite{Affleck:1985wb}, the $\Integer_2$
symmetry is only present in the lattice model if the Hamiltonian is
translation invariant. If it is broken, the transition from the
Haldane phase to the dimerized phase is actually described by means of
a $\text{SU(2)}_1$ WZW model.

In our present paper we deal with a translation invariant SU(N) spin
chain with physical spins in the adjoint representation. Since the
associated Young tableau has $N$ boxes, the central subgroup
$\Integer_N\subset\text{SU(N)}$ is actually acting trivially and the
actual symmetry group is $\text{PSU(N)}=\text{SU(N)}/\Integer_N$ (see
\onlinecite{Duivenvoorden:2012arXiv1206.2462D}). This means that the
group $\Integer_N$ should be anomaly free in the SU(N) WZW model
describing potential phase transitions. A careful inspection of these
models shows that this is the case for all $k$ if $N$ is odd while $k$
is required to be even if $N$ is
even.\cite{Felder:1988CMaPh.117..127F,Lecheminant:2015arXiv150901680L}
We are immediately led to the conjecture that the 2nd order phase
transitions in the phase diagram of our model are generically
described by $\text{SU(N)}_1$ (for odd $N$) and $\text{SU(N)}_2$ (for
even $N$), respectively. Larger values of $k$ can be imagined if the
microscopic system by chance does not feature certain relevant
operators. The latter can not be excluded solely based on symmetry
considerations. The same kind of argument also predicts a value of
$k=2$ for topological phase transitions in translation invariant SU(N)
spin systems which are based on the self-conjugate representation of
Ref.~\onlinecite{Nonne:2012arXiv1210.2072N,Bois:2015PhRvB..91g5121B,Tanimoto:2015arXiv150807601T,Quella:2015}
since the definition of the latter requires $N$ to be even.

Of course, the idea just presented may be generalized to other
continuous symmetry groups~$G$ beyond SU(N). In order to fully
appreciate the generality of our claims it is important to note that
the existence of symmetry protected topological phases crucially
relies on (part of) the central subgroup of~$G$ acting trivially on
the physical representation.\cite{Duivenvoorden:2012arXiv1206.2462D}
It is remarkable that the mathematical structures governing potential
anomalies in critical theories with continuous symmetry group and the
classification of symmetry protected topological phases are related in
such a deep fashion.

Let us finally reiterate that considerations very similar to the ones
above have already appeared before in the
literature.\cite{Affleck:1988NuPhB.305..582A,Furuya:2015arXiv150307292F,Lecheminant:2015arXiv150901680L}
The focus of these works, however, was the stability of gapless
phases. To our knowledge the connection to gapped symmetry protected
topological phases and (topological) phase transitions has not been
stressed so far.

\subsection{Spontaneous inversion symmetry breaking}

In order to obtain unique ground states we deliberately chose to add
inversion symmetry breaking terms to our AKLT Hamiltonians. Of course
one could also simply follow the alternative recipe of
Refs.~\onlinecite{Greiter:PhysRevB.75.184441,Rachel:2010JPhCS.200b2049R,Morimoto:2014PhRvB..90w5111M}
where inversion symmetry in the Hamiltonian is restored at the cost of
having a two-fold degenerate ground state. The corresponding
Hamiltonians should be of the form
\begin{equation}
\begin{split}
  h&=\idop-(\bP_{\bullet}+\bP_S+\bP_A)\\[2mm]
   &=\idop
     -3\gamma N(5N^2-4)\bigl[4\,\vec{S}_1\cdot\vec{S}_2-(\vec{S}_1\cdot\vec{S}_2)^3\bigr]\\[2mm]
   &\qquad-\gamma(7N^2-4)\bigl[4\,(\vec{S}_1\cdot\vec{S}_2)^2-(\vec{S}_1\cdot\vec{S}_2)^4\bigr]
\end{split}
\end{equation}
for an auxiliary spin in the (anti-)fundamental representation and,
similarly,
\begin{equation}
\begin{split}
  h&=\idop-(\bP_{\bullet}+\bP_S+\bP_A+\bP_{S_2})\\[2mm]
   &=\idop
    +4N\delta(N^3+6 N^2+18 N+12)\,\vec{S}_1\cdot\vec{S}_2\\[2mm]
   &\qquad-2\delta(N^4+3 N^3-12 N-8)\,(\vec{S}_1\cdot\vec{S}_2)^2\\[2mm]
   &\qquad-\delta N(3N^2+16N+12)\,(\vec{S}_1\cdot\vec{S}_2)^3\\[2mm]
   &\qquad-\delta(N^2+6 N+4)\,(\vec{S}_1\cdot\vec{S}_2)^4
\end{split}
\end{equation}
for an auxiliary spin in the anti-symmetric representation. Here, the
constants $\gamma$ and $\delta$ are defined by
\begin{equation}
\begin{split}
  \gamma
  &=1/\bigl[8N^2(N^2-1)(N^2-4)\bigr]\quad\text{ and}\\[2mm]
  \delta
  &=1/\bigl[16N^2(N+1)(N+2)\bigr]\;.
\end{split}
\end{equation}
Hamiltonians with lower order in $\vec{S}_1\cdot\vec{S}_2$ can be
obtained by simply projecting out the unwanted contributions without
paying attention to the relative normalization of the remaining
projectors. While the lower order variant of the first expression has
been known for a quite a
while,\cite{Greiter:PhysRevB.75.184441,Rachel:2010JPhCS.200b2049R,Morimoto:2014PhRvB..90w5111M}
the second Hamiltonian (and also the associated lower order variant)
is new, at least to the best of our knowledge.

\section{\label{sc:ParentHamiltonian}A universal parent Hamiltonian}

We now present a general strategy for the construction of a parent
Hamiltonian for an AKLT state. The auxiliary spins $\cV$ and
$\cV^\ast$ as well as the physical spin $\cP$ are assumed to be
arbitrary until further notice as long as the conditions of
Section~\ref{sc:HaldaneGeneral} are satisfied. After outlining the
general case we specialize to the case of interest for us.

\subsection{General strategy}

Our basic idea is to start with a natural Hamiltonian on the
auxiliary level which is then projected onto the physical
level. More precisely, the fundamental object is the projector
\begin{align}
h_{\text{aux}}
=\idop - \ \frac{1}{d(\cV)}
\raisebox{-1mm}{$\includediagram{1.5cm}{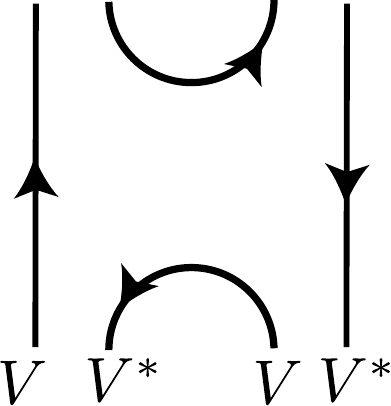}$}
\end{align}
on $(\cV\otimes\cV^\ast)\otimes(\cV\otimes\cV^\ast)$, where $d(\cV)$
denotes the dimension of the space $\cV$. Looking at
Figure~\ref{fig:VBS}, the AKLT state on four auxiliary (two physical)
sites has the spins in the middle joined into a singlet, while the
boundary spins are free. Hence it is clear that it is annihilated by
the above operator.

The auxiliary Hamiltonian needs to be projected onto the physical
subspace space
$\cP\otimes\cP\subset(\cV\otimes\cV^\ast)\otimes(\cV\otimes\cV^\ast)$
using
\begin{align}
\bP_{\text{phys}}
=\raisebox{-2mm}{$\includediagram{2cm}{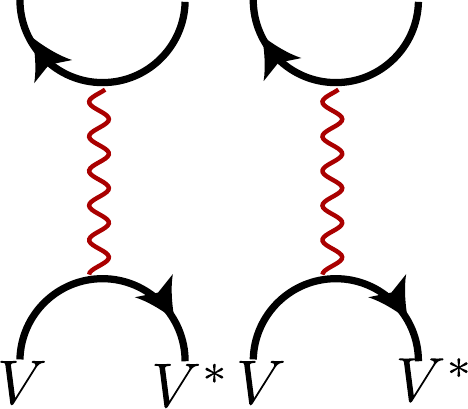}$}\;.
\end{align}
This projection results in the operator
\begin{align}
\bP_{\text{phys}}h_{\text{aux}}\bP_{\text{phys}}
=\ \idop_{\text{phys}}-\raisebox{-.5mm}{$\includediagram{2cm}{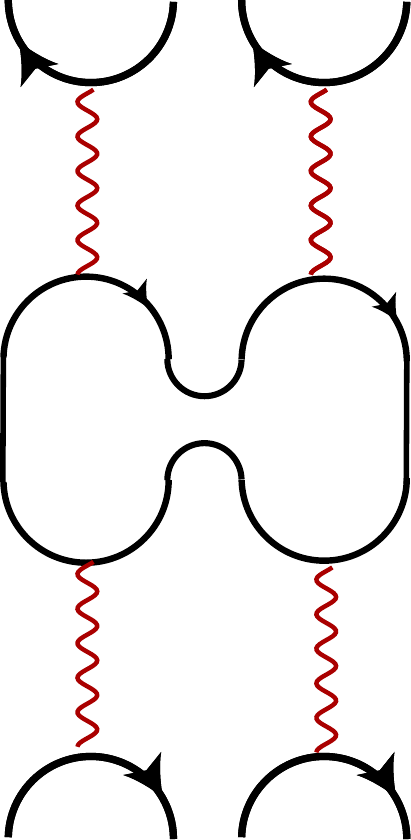}$}\;.
\end{align}
Of course, what we really need is an operator which acts on the
physical two-site Hilbert space $\cP\otimes\cP$. Simply removing the
fringes, gives
\begin{align}
\label{eq:UniversalAKLT}
\tilde{h}_{\text{AKLT}}
=\idop-\ \frac{1}{d(\cV)}\raisebox{-.5mm}{$\includediagram{2cm}{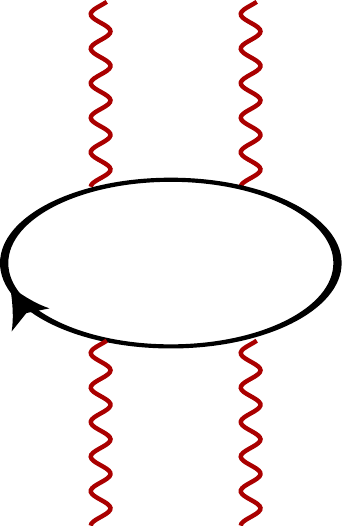}$}\;.
\end{align}
Our central claim is that the above construction provides a
\textbf{general form} of the AKLT Hamiltonian on two physical
sites. Note that \eqref{eq:UniversalAKLT} only requires knowledge of
the `gluon vertex' $\includediagram{.35cm}{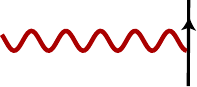}$, which
encodes how the physical spin can be written in terms of the boundary
spins.

Let us discuss the properties of the operator
\eqref{eq:UniversalAKLT}. First, we note that
$\tilde{h}_{\text{AKLT}}$ as defined in \eqref{eq:UniversalAKLT} is
not necessarily a projector. This property, usually deemed necessary,
of AKLT Hamiltonians may indeed be lost during the embedding of the
physical space $\cP$ into $\cV\otimes\cV^\ast$. To be precise,
$\tilde{h}_{\text{AKLT}}$ is a sum of projectors into the constituent
representations of $\cP\otimes\cP$, with non-negative but (generally)
distinct coefficients. Nevertheless, taken on a finite chain we find
the same ground state which along with a gap, indicates that we are in
the same topological phase. A way of `renormalizing' the projectors'
coefficients back to their standard value~$1$ will be outlined after
the next paragraph.

Furthermore, the Hamiltonian \eqref{eq:UniversalAKLT} is generally
{\it not self-conjugate}. Conjugation amounts to reversing each arrow
and physically leads to a parent Hamiltonian for another AKLT state
which can be thought of as the image under inversion or
time-reversal. Only if the auxiliary representation $\cV=\cV^\ast$ is
self-dual, will the associated AKLT state be invariant under these
transformations. This is the case for example, in the original AKLT
state for SU(2) and for the AKLT states suggested in
Ref.~\onlinecite{Nonne:2012arXiv1210.2072N}. In contrast, the
Hamiltonians considered in the present paper are all chiral, with the
sole exception of SU(4), as discussed below.

As the bulk of the paper deals with the Hamiltonians written in terms
of projectors we do the same for \eqref{eq:UniversalAKLT}. As before,
the Hilbert space for two physical sites as
\begin{align}
\cP\otimes\cP
=(\cP_1\oplus \cP_2\oplus\cdots)\oplus\text{rest}\;,
\end{align}
where we distinguish the irreducible representation that are also in
$\cV\otimes\cV^\ast$ from the `rest'. If the $\cP_i$ were distinct
then we could write
\begin{align}
\label{eq:akltproj}
\tilde{h}_{\text{AKLT}} =\idop-(c_1 \mathbb{P}_1 +c_2
\mathbb{P}_2+\cdots)
\end{align}
and normalizing each of the $c_i$ to $1$ gives a true projector if
needed. Indeed, this is nothing but the usual prescription of
projecting out the boundary spins from the space of two physical
spins.

However, for the cases that are of chief interest in our paper, i.e.\
when $\cP$ is the adjoint representation, we have a non-trivial
multiplicity and the right hand side of \eqref{eq:akltproj} is a block
matrix where the diagonal contains projectors and the off-diagonal
entries permute different copies of the degenerate
representation. These coefficients may be computed using
\eqref{eq:UniversalAKLT} and some diagrams, to which we now turn.

\subsection{Computation for the fundamental representation}

For the rest of this subsection
$\cP=\includediagram{.2cm}{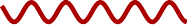}$ will always be the
adjoint and $\cV=\includediagram{.2cm}{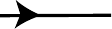}$, the
fundamental representation, with dimensions $N^2-1$ and $N$
respectively. We have
\begin{equation}
\begin{split}
  \cV\otimes\cV &= 0\oplus \cP\\[2mm]
  \cP\otimes\cP &= \cP_{\bullet} \oplus \cP_S \oplus \cP_A + \text{rest}\;.
\end{split}
\end{equation}
Here $\cP_S$ and $\cP_A$ are two copies of the adjoint which are
distinguished by their symmetry under permutation. Diagrammatically,
\begin{align}
\raisebox{-.5mm}{$\includediagram{1.5cm}{figs/appeq15.pdf}$}
=\underbrace{\frac{1}{N^2-1}\,\raisebox{-.5mm}{$\includediagram{1.5cm}{figs/adjtrace.pdf}$}}_{\bP_\bullet}
\quad+\quad\underbrace{\frac{1}{2N}\raisebox{-.5mm}{$\includediagram{1.5cm}{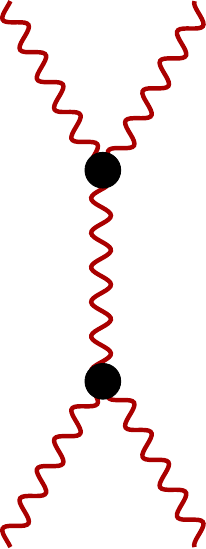}$}}_{\bP_A}\\
\quad+\quad\underbrace{\frac{N}{2(N^2-4)}\raisebox{-.5mm}{$\includediagram{1.5cm}{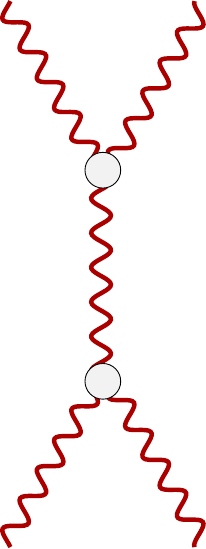}$}}_{\bP_S}
\ +\quad\text{rest} \label{eq:twoadj1}\;,
\end{align}
where we use the expressions listed in Appendix~\ref{sc:Projectors}.
The Hamiltonian from \eqref{eq:UniversalAKLT} may be decomposed as,
\begin{align}
\tilde{h}_{\text{AKLT}} = \idop - (c_\bullet\bP_\bullet +c_S\bP_S +c_A\bP_A + c_{AS}\bX). \label{eq:akltfun}
\end{align}
It is convenient to write things in terms of a block matrix in the
$\cP_S\oplus\cP_A$ space. Then $\Pi=\smat1&0\\0&-1\stam$ is the parity
operator, and $\bX$ is defined (up to sign) by the requirements that
$\bX\Pi=-\Pi\bX$ and $\bX^2=\smat 1&0\\0&1\stam$. We may then collect
the last three terms of \eqref{eq:akltfun} into the symmetric matrix
\begin{align}
M=\mat c_S& c_{AS}\\c_{AS}&c_A\tam
\quad \text{ with }\quad
M^2 = \alpha M\;.\label{eq:Mproj}
\end{align}
In other words, $M$ is proportional to a projector. Its diagonal
entries can be extracted by sandwiching $\tilde{h}_{\text{AKLT}}$
within the appropriate projectors. To illustrate this procedure, we
show the computation of $c_A$. From \eqref{eq:akltfun},
$\bP_A\,(\idop-\tilde{h}_{\text{AKLT}})\, \bP_A = c_A\,\bP_A$ and from
\eqref{eq:UniversalAKLT} and \eqref{eq:twoadj1}
\begin{align}
(2N\bP_A)\bigl[N (\idop-\tilde{h}_{\text{AKLT}})\bigr](2N\bP_A)=
\includediagram{2.5cm}{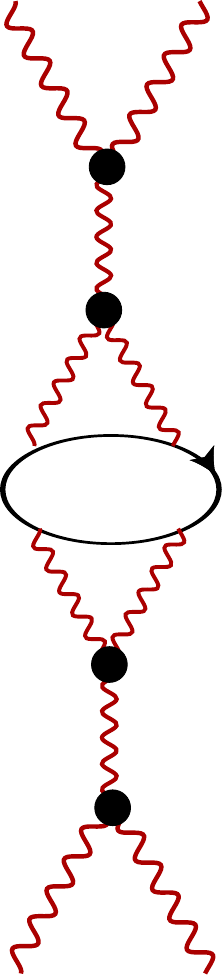}
=\includediagram{2.5cm}{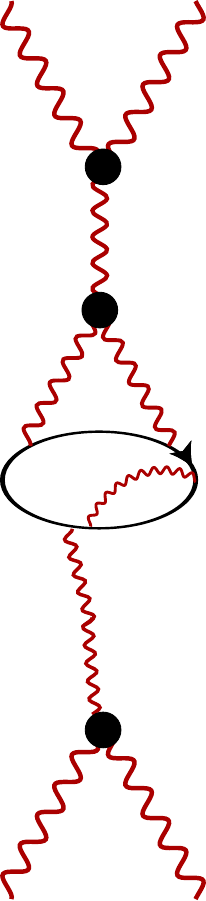}
-\includediagram{2.5cm}{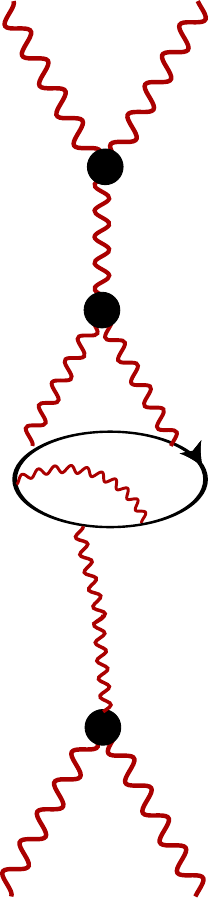}\\
=\left(-N+\frac1N-\frac1N\right)
\includediagram{2.5cm}{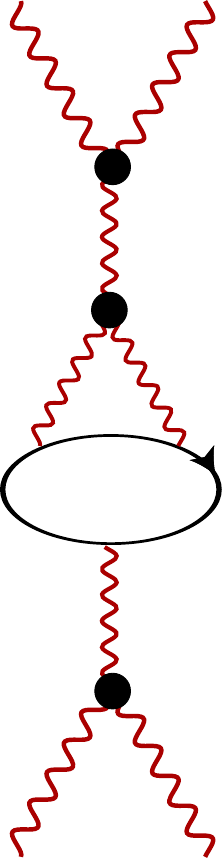}
=N^2
\includediagram{2.5cm}{figs/adjA.pdf}
=N^3\bP_A,
\end{align}
where we have repeatedly used the Lie algebra identity \eqref{eq:FundId2}. 
So $\bP_A\, \tilde{h}_{\text{AKLT}}\, \bP_A = 1/2\,\bP_A$ and $c_A = \frac12$.

A similar calculation gives $c_S$ (which can also be read off from
Table~\ref{tab:Eigenvalues}) and the off-diagonal entry $c_{AS}$ is
worked out from the condition in \eqref{eq:Mproj}. Finally we divide
by $\alpha$ and set $c_\bullet =1$ to get the normalized Hamiltonians
displayed in Eqs.~\eqref{eq:AKLT1-1}.

\subsection{Computation for the antisymmetric representation}

We now consider the case
$\cV=\includediagram{.35cm}{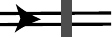}=\tinyyng{1,1}$ with
dimension $N(N-1)/2$. Recall that the dark bars in the diagram stand
for anti-symmetrization. From
\begin{equation}
\begin{split}
\cV\otimes\cV^\ast
&=0\oplus\cP
\oplus\cP_{S_2}\\[2mm]
\cP\otimes\cP
&=\cP_{\bullet}
\oplus\cP_A
\oplus\cP_S
\oplus\cP_{S_2}
\oplus\text{rest}
\end{split}
\end{equation}
we see that we have to deal with an additional projector
$\bP_{S_2}$. From the universal form~\eqref{eq:UniversalAKLT},
\begin{align}
\tilde{h}_{\text{AKLT}}
=\idop-\frac{2}{N(N-1)}\includediagram{2cm}{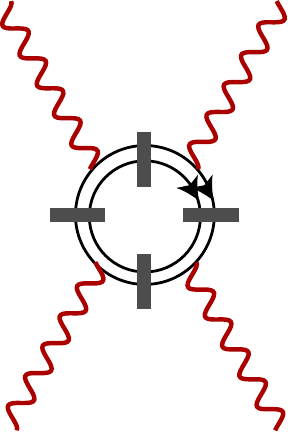}\;,
\end{align}
we find after some bookkeeping that
\begin{widetext}
	\begin{align}
	\includediagram{2cm}{figs/appeq22.pdf} \quad=&\quad
	(N-4)\includediagram{2cm}{figs/appeq23.pdf}\quad+\quad
	\includediagram{2cm}{figs/adjtrace.pdf}\quad+\quad
	\includediagram{2cm}{figs/appeq15.pdf}\quad+\quad
	\includediagram{2cm}{figs/appeq16.pdf}\\
	&-\left\{\includediagram{2cm}{figs/appeq24.pdf}+
	\includediagram{2cm}{figs/appeq25.pdf}
	\right\}\quad-\quad
	\left\{\includediagram{2cm}{figs/appeq26.pdf}+
	\includediagram{2cm}{figs/appeq27.pdf}
	\right\}\;.
	\end{align}
\end{widetext}
Note that for $N=4$ the first term on the right disappears and the
rest is clearly invariant both under inversion, as well as
conjugation (arrow reversal). This is the inversion symmetric
topological phase for SU(4) that has been considered previously in
Section~\ref{sc:SU4}. In terms of projectors,
\begin{align}
\tilde{h}_{\text{AKLT}} = \idop - (c_\bullet\bP_\bullet +c_A\bP_A +c_S\bP_S + c_{AS}\bX +\bP_{S_2})\;. \label{eq:akltfun2}
\end{align}
$c_A$ and $c_S$ may be worked out from Table~\ref{tab:Eigenvalues} and
applying the normalization as shown before, gives the results in
Eq.~\eqref{eq:AKLT2-1}.

\section{\label{sc:Conjecture}Comments on Haldane's Conjecture}

The famous Haldane Conjecture for SU(2) states that anti-ferromagnetic
Heisenberg spin chains behave differently, depending on whether the
spin $s$ is an integer or half-integer. In the former case, assuming
isotropy and in the limit of infinite length, we have a gapped phase
with a unique ground state while the latter is gapless. This
conjecture is well-supported by numerical evidence and may be arguable
considered proven, in the large $s$
limit.\cite{Affleck:1987PhRvB..36.5291A,Affleck:1989JPCM....1.3047A}

For SU(N), there are two obvious directions in which to
generalize. First, we may ask which representations realize a gapless
phase. One result by Affleck and Lieb \cite{Affleck:1986pq} states
that rectangular Young tableaux where the total number of boxes has no
common factors with $N$, are gapless. Greiter and Rachel suggested to
extend this to non-rectangular tableau with the same divisibility
condition.\cite{Greiter:PhysRevB.75.184441} Here we would like to
point out that this non-divisibility condition exactly {\it rules out}
a non-trivial topological phase which is protected by any of the
groups $\text{SU(N)}/\Integer_m$ where $m$ divides
$N$.\cite{Duivenvoorden:2012arXiv1206.2462D} In other words, from
symmetry considerations alone, the phase must either be gapless or
topologically trivial.

The second direction concerns the gapped phases -- for simplicity we
restrict ourselves to the case where $N$ divides the number of boxes
in the tableau. Assuming the Heisenberg model is indeed gapped, what
is the topological class (the protected boundary spins if any)? For
SU(2) it is well known that the spin $1$ chain has protected spin
$1/2$ modes at the boundary. Since the Heisenberg Hamiltonian is
inversion symmetric, a unique ground state implies the boundary spin
to be self-conjugate too. This is indeed possible for SU(4) if the
boundary states are in a six-dimensional $\cV=\tinyyng{1,1}$
multiplet.
\medskip

{\bf Conjecture}: The Heisenberg chain for SU(4) in the adjoint
representation is in the same phase as the AKLT Hamiltonian of
Eq.~\eqref{eq:AKLT2-2}.
\medskip

When the physical spin is in the representation $\tinyyng{2,2}$, the
analogous statement has been convincingly shown in
Refs.~\onlinecite{Bois:2015PhRvB..91g5121B,Tanimoto:2015arXiv150807601T,Quella:2015}.

Finally we turn to the chiral phases which break inversion symmetry,
and occupy most of this paper. A natural question is whether the
Heisenberg model flows to such a chiral gapped phase when perturbed by
a small inversion breaking term e.g.\ $\bC_A$. Both this question and
the above conjecture are within the scope of present DMRG numerical
methods.

\section{Conclusions and Outlook}

In this paper we have provided a systematic analysis of Haldane phases
in SU(N) spin chains with physical spins transforming in the adjoint
representation. With the help of birdtracks we succeeded in
constructing AKLT Hamiltonians for auxiliary spins transforming in the
fundamental $\tinyyng{1}$ or the rank-2 anti-symmetric representation
$\tinyyng{1,1}$. It should be emphasized that our Hamiltonians have
{\em unique} ground states which, by construction, reside in
topological sectors characterized by the $\Integer_N$ quantum numbers
$\pm[1]$ and $\pm[2]$.\cite{Duivenvoorden:2012arXiv1206.2462D} This is
in stark contrast to earlier
investigations\cite{Affleck:1991NuPhB.366..467A,Greiter:PhysRevB.75.184441,Rachel:2010JPhCS.200b2049R,Morimoto:2014PhRvB..90w5111M}
where only Hamiltonians leading to a two-fold ground state degeneracy
and an associated spontaneous inversion symmetry breaking have been
considered. For SU(4) our analysis gives the first complete account of
all existing Haldane phases.\cite{Duivenvoorden:2012arXiv1206.2462D}

On the way we had to overcome a number of technical complications. The
main problem arose from the fact that the decomposition of the
physical two-site Hilbert space into irreducible representations of
SU(N) features a non-trivial multiplicity. In this space we had to
identify a specific one-dimensional subspace and we showed how this
can be achieved transparently using birdtracks. Besides helping us
with the construction of the AKLT states and their associated
Hamiltonians this method also gave access to the eigenvalues of the
ground states' transfer matrix. As a consequence, we have been able to
confirm the existence of a spectral gap, partially reproducing earlier
results in the literature. It should be clear that our method can
easily be extended to the determination of spin-spin correlation
functions (see Ref.~\onlinecite{Morimoto:2014PhRvB..90w5111M} for an
algebraic derivation for the first of our two cases).

In a first and rather sketchy attempt to characterize the phase
diagram of our spin chains we focused on the potential presence of
special integrable and/or critical points. As one of the remarkable
features the phase diagram shows the existence of an integrable point
with enhanced $\text{SU($N^2-1$)}$ symmetry and an effective
description in terms of a $\text{SU($N^2-1$)}$ WZW model at level
$k=1$. At least one other integrable point is likely to exist but the
explicit construction of its Hamiltonian requires the fusion procedure
for integrable
models\cite{Andrei:1984PhLA..104..370A,Johannesson:1986NuPhB.270..235J}
and has not been attempted here.

Let us stress that our paper also contains two results which are of
rather general importance beyond the specific setup we have been
investigating. The first one concerns a proposal for a universal
parent Hamiltonian in Section~\ref{sc:ParentHamiltonian} which may
function as a convenient replacement for the standard AKLT
construction, also in the case of other symmetry groups and/or
representations.

Secondly, we found an intimate relation between the classification of
symmetry protected topological phases with continuous symmetry
groups\cite{Duivenvoorden:2012arXiv1206.2462D} and a potential
multicriticality of topological phase transitions. Specifically, we
have been led to conjecture that the critical points in the phase
diagram of our model for even values of $N$ should all be described by
SU(N) WZW models with level $k=2$ (except at fine-tuned points where
$k$ may be even larger). This, as well as our conjecture about the
nature of the phase realized in the SU(4) Heisenberg chain, certainly
deserve future study.

Of course there are also a few important issues that we did not touch
upon in the present article. One concerns the nature of the
excitations above the ground state which may lead to a similar
interpretation as in Ref.~\onlinecite{Greiter:PhysRevB.75.184441}. It
is expected that our precise knowledge of the Hamiltonians for general
values of $N$ will facilitate future large-$N$ analyses. Another
direction of research would be to enlarge the phase diagram by
admitting terms in the Hamiltonian which break the SU(N) symmetry but
still preserve the subgroup
$\Integer_N\times\Integer_N\subset\text{PSU(N)}$. As was discussed in
Ref.~\onlinecite{Duivenvoorden:2013arXiv1304.7234D} (see also
Ref.~\onlinecite{Else:2013arXiv1304.0783E}) this symmetry protects the
same type of Haldane phases which are hence stable against such
deformations. Some studies in this direction have been performed in
Refs.~\onlinecite{Morimoto:2014PhRvB..90w5111M,Bois:2015PhRvB..91g5121B,Tanimoto:2015arXiv150807601T},
albeit mostly for $N=3$ or a different physical representation.
Finally, in view of the experimental progress in the realization of
systems exhibiting SU(N) magnetism it is a pressing question to which
extent the phases we have been discussing can actually be realized in
a variant of the Fermi-Hubbard model or even in
experiment.\cite{Gorshkov:2010NatPh...6..289G,Zhang:2014Sci...345.1467Z,Scazza:2014NatPh..10..779S,Pagano:2014NatPh..10..198P,Hofrichter:2015arXiv151107287H}

\begin{acknowledgments}
  We would like to thank Kasper Duivenvoorden, Philippe Lecheminant,
  Frederic Mila, Pierre Nataf, Michael Stone and Martin Zirnbauer for
  useful discussions. We, moreover, gratefully acknowledge the support
  and the stimulating atmosphere during the Workshop ``Topological
  Phases of Quantum Matter'' at the ESI in Vienna where part of this
  work was carried out. Both authors are funded by the German Research
  Foundation (DFG) through Martin Zirnbauer's Leibniz Prize, DFG grant
  no. ZI 513/2-1. Additional support was received by the DFG through
  the SFB$|$TR12 ``Symmetries and Universality in Mesoscopic Systems''
  and the Center of Excellence ``Quantum Matter and Materials''
  (QM$^2$).
\end{acknowledgments}

\appendix
\begin{table*}
	\begin{center}
		\footnotesize
		\begin{tabular}{c|ccccccc}
			$\tinyyng{2,1,1,1,1}\otimes\tinyyoung{11,2,\,,\,,r}$
			& $\bullet$
			& $\tinyyoung{\,1,1,2,3,4}$ 
			& $\tinyyoung{\,1,2,3,4,r}$
			& $\tinyyoung{\,11,2,3,4}$
			& $\tinyyoung{\,\,1,\,12,\,\,,\,\,,\,r}$\rule[-1.7em]{0pt}{1em}
			& $\tinyyoung{\,\,11,\,2,\,\,,\,\,,\,r}$
			& $\tinyyoung{\,1,12,3,4}$\\\hline\hline&&&&&&&\\[-1em]
			Symmetry & S & S & A & A & A & S & S\\
			Projector & $\bP_{\bullet}$ & $\bP_S$ & $\bP_A$ & $\bP_{A_1}$ & $\bP_{A_2}$ & $\bP_{S_1}$ & $\bP_{S_2}$\\
			Dimension & $1$& $N^2{-}1$ & $N^2{-}1$ & $\frac{(N^2-1)(N^2-4)}{4}$ & $\frac{(N^2-1)(N^2-4)}{4}$ & $\frac{N^2(N+3)(N-1)}{4}$ & $\frac{N^2(N-3)(N+1)}{4}$  \\\hline&&&&&&&\\[-1em]
			$\includediagram{1cm}{figs/appeq23.pdf}$ & $\frac{N^2-1}{N}$ & \multicolumn{2}{c}{$\smat\frac{N^2-4}{2N}&\pm\frac{\sqrt{N^2-4}}{2}\\\pm\frac{\sqrt{N^2-4}}{2}&\frac{N}{2}\stam$} & $0$ & $0$ & $0$&$0$\\[2mm]						$\includediagram{1cm}{figs/appeq23.pdf}-\includediagram{1cm}{figs/appeq23a.pdf}$ 
			& $0$ & \multicolumn{2}{c}{$\smat0&\pm\sqrt{N^2-4}\\\pm\sqrt{N^2-4}&0\stam$} & $0$ & $0$ & $0$&$0$\\\hline&&&&&&&\\[-1em]
			$\includediagram{1cm}{figs/appeq26.pdf}$ 
			& $-\frac{1}{N}$ & \multicolumn{2}{c}{$\smat -\frac{2}{N}&0\\0&0\stam$} & $1$ & $-1$ & $1$ & $-1$\\[2mm]						$\includediagram{1cm}{figs/appeq26.pdf}-\includediagram{1cm}{figs/appeq27.pdf}$ 		& $0$ & \multicolumn{2}{c}{$\smat 0&0\\0&0\stam$} & $2$ & $-2$ & $0$ & $0$\\\hline&&&&&&&\\[-1em]
			$\includediagram{1cm}{figs/appeq16.pdf}$  & $1$ &\multicolumn{2}{c}{$\smat 1&0\\0&-1\stam$} & $-1$ & $-1$ & $1$ & $1$ \\[2mm]
			$\includediagram{1cm}{figs/appeq24.pdf}+
			\includediagram{1cm}{figs/appeq25.pdf}$ & $\frac{N^2-1}{N}$ &\multicolumn{2}{c}{$\smat \frac{N^2-4}{N}&0\\0&-N\stam$} & $0$ & $0$ & $0$ & $0$
		\end{tabular}
		\caption{\label{tab:Eigenvalues}Eigenvalues of some tensors, represented as diagrams, including all the results used in the text.}
	\end{center}
\end{table*}

\begin{widetext}
\bigskip
\section{\label{sc:Projectors}Diagrammatic form of the projectors}
For convenience we include the diagrammatic form of the projectors
into the irreducible representations inside product of two adjoint
representations,
\begin{align}
  \includediagram{1.5cm}{figs/appeq15.pdf}
  =\bP_{\bullet}+\bP_{A}+\bP_{S}+\bP_{A_1}+\bP_{A_2}+\bP_{S_1}+\bP_{S_2}
\end{align}
where\cite{Cvitanovic}
\begin{align}
  \bP_{\bullet}
  &=\frac{1}{N^2-1}\includediagram{1.5cm}{figs/adjtrace.pdf}&
  \bP_{A}
  &=\frac{1}{2N}\includediagram{1.5cm}{figs/adjA.pdf}&
  \bP_{S}
  &=\frac{N}{2(N^2-4)}\includediagram{1.5cm}{figs/adjS.pdf}
\end{align}
\begin{align}
  \bP_{A_1}
  &=\frac12\,\left\{\includediagram{1.5cm}{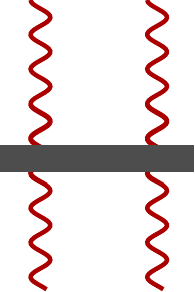}
    +\frac{1}{2N}\includediagram{1.5cm}{figs/adjA.pdf}
    +\includediagram{1.5cm}{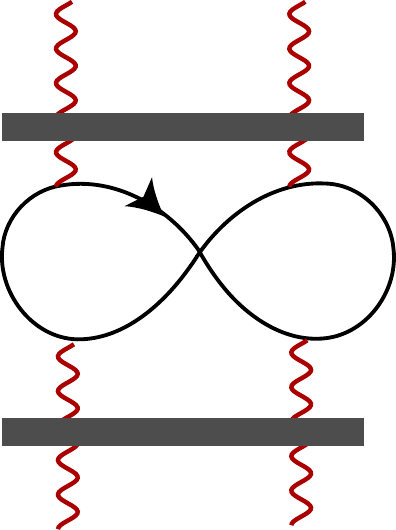}\right\}&
  \bP_{A_2}
  &=\frac12\,\left\{\includediagram{1.5cm}{figs/appeq14.pdf}
    +\frac{1}{2N}\includediagram{1.5cm}{figs/adjA.pdf}
    -\includediagram{1.5cm}{figs/appeq18.pdf}\right\}
\end{align}
\begin{align}
  \bP_{S_1}
  &=\frac12\,\left\{\includediagram{1.5cm}{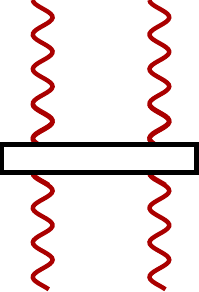}
    +\frac{1}{2(N-2)}\includediagram{1.5cm}{figs/adjS.pdf}
    -\includediagram{1.5cm}{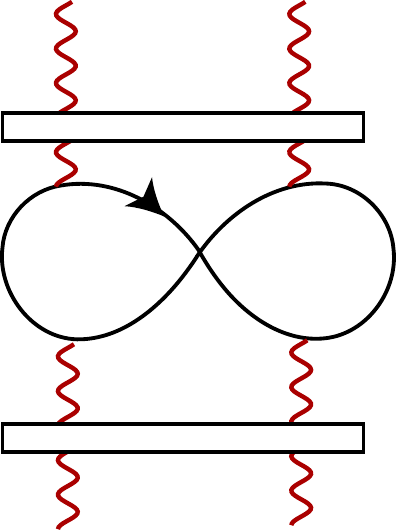}
    -\frac{1}{N(N-1)}\includediagram{1.5cm}{figs/adjtrace.pdf}\right\}\\
  \bP_{S_2}
  &=\frac12\,\left\{\includediagram{1.5cm}{figs/appeq13.pdf}
    +\frac{1}{2(N+2)}\includediagram{1.5cm}{figs/adjS.pdf}
    +\includediagram{1.5cm}{figs/appeq17.pdf}
    -\frac{1}{N(N-1)}\includediagram{1.5cm}{figs/adjtrace.pdf}\right\}
\end{align}
The projectors can also be expressed in terms of spin operators. This
results in
\begin{align}
  \bP_{\bullet}
  &\ =\ (\bQ-2N)(\bQ-4N)(\bQ-4N-4)(\bQ-4N+4)/\bigl[128N^2(N+1)(N-1)\bigr]\\[2mm]
  \bP_S
  &\ =\ \bK\bQ(\bQ-4N)(\bQ-4N-4)(\bQ-4N+4)/\bigl[128N^4(N+2)(N-2)\bigr]\\[2mm]
  \bP_A
  &\ =\ -(\bK+8N^2)\bQ(\bQ-4N)(\bQ-4N-4)(\bQ-4N+4)/\bigl[128N^4(N+2)(N-2)\bigr]\\[2mm]
  \bP_{A_1/A_2}
  &\ =\ -(\bC_S+4N)\bQ(\bQ-2N)(\bQ-4N-4)(\bQ-4N+4)/\bigl[1024N^3\bigr]\\[2mm]
  \bP_{A_2/A_1}
  &\ =\ (\bC_S-4N)\bQ(\bQ-2N)(\bQ-4N-4)(\bQ-4N+4)/\bigl[1024N^3\bigr]\\[2mm]
  \bP_{S_1}
  &\ =\ \bQ(\bQ-2N)(\bQ-4N)(\bQ-4N+4)/\bigl[256(N+1)(N+2)\bigr]\\[2mm]
  \bP_{S_2}
  &\ =\ \bQ(\bQ-2N)(\bQ-4N)(\bQ-4N-4)/\bigl[256(N-1)(N-2)\bigr]\ \ .
\end{align}
We outline the proof of the relation \eqref{eq:XCA} obeyed by the
operator $\bC_A$ since it is quite important. First, we note that
$\bC_A$ is odd under inversion and hence must take an irreducible
representation to another of opposite parity. Next we compute,
\begin{align*}
\left(\frac{\bC_A}{2N}\right)^2 &= \left(\includediagram{1.5cm}{figs/appeq23.pdf}-\includediagram{1.5cm}{figs/appeq23a.pdf}\right)^2 = 
\frac{N}{2}\includediagram{1.5cm}{figs/adjS.pdf}+(N^2-4)\includediagram{1.5cm}{figs/adjA.pdf}
=(N^2-4)(\bP_S+\bP_A)
\end{align*}
which proves \eqref{eq:XCA} (the second equality follows from a long
diagrammatic manipulation).
\end{widetext}


\begin{thebibliography}{10}
\bibitem{Quella:2015}
T.~Quella and A.~Weichselbaum, ``{Haldane phases of $SU(N)$ spin chains}.''
  Work in progress (some of the results of these investigations have already
  been presented on several conferences and workshops).
\end{thebibliography}

\def\cprime{$'$}
\providecommand{\href}[2]{#2}\begingroup\raggedright\endgroup

\end{document}